\documentclass[10pt,twocolumn,letterpaper]{article}

\usepackage{iccv}
\usepackage{times}
\usepackage{epsfig}
\usepackage{graphicx}
\usepackage{amsmath}
\usepackage{amssymb}
\usepackage{booktabs}
\usepackage{pifont}
\usepackage{setspace}
\usepackage{algpseudocode}
\usepackage[accsupp]{axessibility}

\usepackage[pagebackref=true,breaklinks=true,colorlinks,bookmarks=false]{hyperref}

\iccvfinalcopy % *** Uncomment this line for the final submission

\def\iccvPaperID{2526} % *** Enter the ICCV Paper ID here

\usepackage[capitalize]{cleveref}
\crefname{section}{Sec.}{Secs.}
\Crefname{section}{Section}{Sections}
\Crefname{table}{Table}{Tables}
\crefname{table}{Tab.}{Tabs.}

\usepackage{bbding}
\usepackage{soul}
\usepackage{bibunits}
\usepackage{bm}
\usepackage{array}
\usepackage{xcolor}
\usepackage{pdflscape}
\usepackage{makecell}
\usepackage{framed,multirow}
\usepackage[flushleft]{threeparttable}
\usepackage{amssymb}% http://ctan.org/pkg/amssymb
\usepackage{pifont}% http://ctan.org/pkg/pifont
\usepackage{algorithm}
\usepackage{layouts}
\usepackage{listings}
\usepackage{pythonhighlight}
\makeatletter
\newcommand\footnoteref[1]{\protected@xdef\@thefnmark{\ref{#1}}\@footnotemark}
\makeatother

\newcolumntype{P}[1]{>{\centering\arraybackslash}p{#1}}
\newlength\savewidth\newcommand\shline{\noalign{\global\savewidth\arrayrulewidth
  \global\arrayrulewidth 0.8pt}\hline\noalign{\global\arrayrulewidth\savewidth}}

\definecolor{Highlight}{HTML}{39b54a}  % green

\newenvironment{jlred}{\color{red}}{}
\newenvironment{jlgreen}{\color{green}}{}
\newenvironment{jlblue}{\color{blue}}{}

\begin{document}

\title{CLIP-Driven Universal Model for Organ Segmentation and Tumor Detection}

\author{Jie~Liu$^1$, Yixiao~Zhang$^2$, Jie-Neng~Chen$^2$, Junfei~Xiao$^2$, Yongyi~Lu$^2$, Bennett A. Landman$^3$,\\Yixuan Yuan$^{4,5}$, Alan~Yuille$^2$, Yucheng~Tang$^{3,6,*}$, and Zongwei~Zhou$^{2,}$\thanks{Corresponding authors: Yucheng~Tang (\href{mailto:yuchengt@nvidia.com}{yuchengt@nvidia.com}) and Zongwei~Zhou (\href{mailto:zzhou82@jh.edu}{zzhou82@jh.edu})}\\[2mm]
$^1$City University of Hong Kong~~~~$^2$Johns Hopkins University~~~~$^3$Vanderbilt University \\ $^4$Chinese University of Hong Kong~~~$^5$CUHK Shenzhen Research Institute~~~$^6$NVIDIA \\ [0.5mm]
{\small Project:~\href{https://github.com/ljwztc/CLIP-Driven-Universal-Model}{https://github.com/ljwztc/CLIP-Driven-Universal-Model}}
}

\maketitle

\thispagestyle{empty}

%%%%%%%%% ABSTRACT
\begin{abstract}

An increasing number of public datasets have shown a marked impact on automated organ segmentation and tumor detection. However, due to the small size and partially labeled problem of each dataset, as well as a limited investigation of diverse types of tumors, the resulting models are often limited to segmenting specific organs/tumors and ignore the semantics of anatomical structures, nor can they be extended to novel domains. 
To address these issues, we propose the CLIP-Driven Universal Model, which incorporates text embedding learned from Contrastive Language-Image Pre-training (CLIP) to segmentation models. This CLIP-based label encoding captures anatomical relationships, enabling the model to learn a structured feature embedding and segment 25 organs and 6 types of tumors. The proposed model is developed from an assembly of 14 datasets, using a total of 3,410 CT scans for training and then evaluated on 6,162 external CT scans from 3 additional datasets. We rank first on the Medical Segmentation Decathlon (MSD) public leaderboard and achieve state-of-the-art results on Beyond The Cranial Vault (BTCV). Additionally, the Universal Model is computationally more efficient (6$\times$ faster) compared with dataset-specific models, generalized better to CT scans from varying sites, and shows stronger transfer learning performance on novel tasks. 

% Code and models are available at \href{https://anonymous.4open.science/r/clip_anona-65FF/README.md}{anonymous link}.
% The design of CLIP embedding enables the Universal Model to be easily extended to new classes without catastrophically forgetting the previously learned classes.

\end{abstract}

\section{Introduction}
\label{sec:introduction}

\begin{figure}[t]
\centerline{\includegraphics[width=0.59\linewidth]{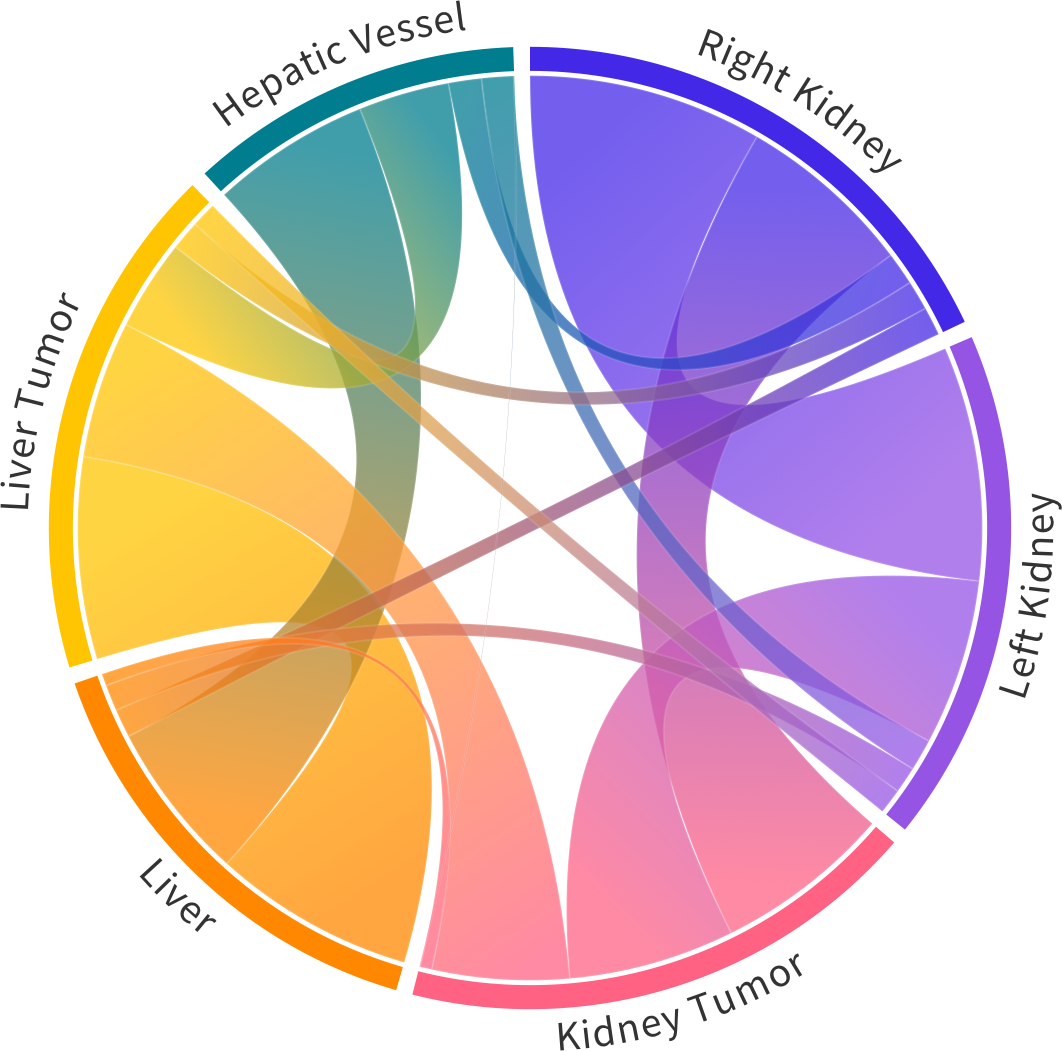}}
\caption{
\textbf{Cosine similarity between CLIP embeddings.} The CLIP embedding reveals the intrinsic semantics of the anatomical structures by mapping similar concepts close to each other in the embedding space. For example, ``Liver'' has a large similarity with ``Liver Tumor'' and ``Hepatic Vessel'' (the hepatic vessel returns low-oxygen blood from the liver to the heart, which has a high anatomical relationship with the liver).
% ``Left Kidney'' has a large similarity with ``Right Kidney''.
}
\label{fig:teaser}
\end{figure}

Enormous advances in medical imaging benefit from the ever-growing number of annotated datasets~\cite{ma2021abdomenct,antonelli2021medical,luo2021word,ji2022amos,wasserthal2022totalsegmentator}. Although a total of around 5,000 annotated abdominal CT scans are publicly available, it is still commonly perceived that medical imaging datasets are too small to develop robust AI models~\cite{zhou2021towards,wang2021annotation,piccialli2021survey,tajbakhsh2020embracing,zhou2022interpreting,chen2022recent}. One reason for this impression is the high cost of detailed per-voxel segmentation annotations, which can take nearly one hour per organ for an expert annotator. Since each institute has time, monetary, and clinical constraints, the number of CT scans in each dataset is limited, and the types of annotated organs vary significantly from institute to institute. Moreover, only a small proportion (hundreds) of public CT scans contain tumor annotation performed by experts~\cite{bilic2019liver,heller2020international,antonelli2021medical}.

The partially labeled problem~\cite{kang2021data,zhang2021dodnet,li2019partial} can impose significant limitations on the performance of models trained on existing public datasets, ultimately hindering their effectiveness for multi-organ segmentation and tumor detection. However, despite this challenge, the potential of AI models in these areas remains promising and largely unexplored. 
This has motivated us to exploit the public datasets with partial labels, and demonstrate the clinical impact of AI framework, including model expansibility (\ie, adaptable to various network backbone), generalizability (\ie, robust to CT scans from various hospitals)~\cite{ma2021abdomenct} and transferability (\ie, generic image representation that is transferable to multiple downstream tasks)~\cite{zhou2021models}. Specifically, we have assembled 14 publicly available datasets, including 3,410 CT scans with 25 partially annotated organs and 6 tumors. 

%Furthermore, we anticipate that most of the medical datasets, released in the near future, will also focus on a small set of organs/tumors and that some current unlabeled organs/tumors, such as the vermiform appendix, may be annotated. This requires us to develop new strategies that can continually deal with more partially labeled datasets with novel classes from a variety of institutes. 

Formidable challenges exist in assembling partially annotated datasets. \textbf{First,} label inconsistency, in five aspects.
\textit{(i)} Index inconsistency. The same organ can be labeled as different indexes. For example, the stomach is labeled `7' in BTCV, but `5' in WORD.
\textit{(ii)} Name inconsistency. Naming can be confusing if multiple labels refer to the same anatomical structure. For example, ``postcava'' in AMOS22 and ``inferior vena cava'' in BTCV.
\textit{(iii)} Background inconsistency. For example, when combining Pancreas-CT and MSD-Spleen, the pancreas is marked as the background in MSD-Spleen, but it should have been marked as the foreground.
\textit{(iv)} Organ overlapping. There is overlap between various organs. For example, ``Hepatic Vessel'' is part of the ``Liver'' and ``Kidney Tumor'' is a sub-volume of the ``Kidney''.
\textit{(v)} Data overlapping. Some CT scans are overlapped among public datasets, but with different annotations. For example, KiTS is part of AbdomenCT-1K, and kidney tumor is annotated in KiTS rather than AbdomenCT-1K.
\textbf{Second,} label orthogonality. 
Most segmentation methods, trained with one-hot labels~\cite{zhang2021dodnet}, ignore the semantic relationship between classes.
Given one-hot labels of liver [1,0,0], liver tumor [0,1,0], and pancreas [0,0,1], there is no semantic difference between liver$\leftrightarrow$liver tumor and liver$\leftrightarrow$pancreas. 
A possible solution is few-hot labels~\cite{shi2021marginal}, with which, the liver, liver tumor, and pancreas can be encoded as [1,0,0], [1,1,0], and [0,0,1]. Although few-hot labels could indicate that liver tumors are part of the liver, the relationship between organs remains orthogonal.
% neither one-hot nor few-hot labels are easy to extend to more classes~\cite{he2021incremental,chen2022class}. Adding novel classes requires increasing the dimensionality of one- or few-hot labels and retraining the previously trained model.

To address above mentioned challenged, CLIP-driven \textit{Universal Model} incorporates text embedding and adopts masked back-propagation mechanism with binary segmentation mask. Specifically, we maintain a revised label taxonomy derived from a collection of public datasets and generate a binary segmentation mask for each class during image pre-processing. For architecture design, we draw inspiration from Guo~\etal~\cite{guo2016entity} and replaced one- or few-hot labels with the text embedding generated by the pre-trained text encoder from CLIP\footnote{CLIP (Contrastive Language–Image Pre-training) was pre-trained on 400 million image-text pairs (some are medical images and text~\cite{chambon2022adapting}), exploiting the semantic relationship between images and language.}. \figureautorefname~\ref{fig:teaser} illustrates how CLIP embedding presents the relationship between organs and tumors. This CLIP-based label encoding enhances the anatomical structure of universal model feature embedding, which is visualized in \figureautorefname~\ref{fig:tsne}. At last, we only compute loss for the classes with available labels.
% More importantly, the fixed-length CLIP embedding allows us to adapt the pre-trained model to open-vocabulary segmentation and to extend to novel classes without the loss of previously learned classes.

In summary, this work proposes a CLIP-Driven \textit{Universal Model} that allows superior segmentation of 25 organs and detection of 6 tumors with state-of-the-art performance. The Universal Model can be generalized to CT scans from different institutes. Experimental results have demonstrated \textbf{six advantages} of the CLIP-Driven Universal Model:
\begin{enumerate}
    \item High abdominal organ segmentation performance. We rank first in the MSD and BTCV challenges, leading to substantial performance improvement. Moreover, six organs can be annotated by Universal Model with a similar intra-observer variability to humans.
    
    \item Predicting fewer false positives than existing models while maintaining high sensitivity for tumor detection.

    \item Computationally more efficient than dataset-specific models, accelerating the testing speed by factor of six.

    \item The Universal Model framework can be expanded to various backbones such as CNNs and Transformers.

    \item The performance of organ segmentation and tumor detection can be generalized to CT scans from a variety of hospitals without additional tuning and adaptation.

    \item An effective Foundation Model for numerous downstream tasks, showing a strong transferability on tasks across multiple diseases, organs, and datasets.
    
\end{enumerate}

% {\jlred With the Universal Model, we have also created a large dataset of 3,672 CT scans with 6 organs annotated by either experts or the model.} Refinement of model prediction for some cases is performed. This dataset, comprising multi-center, multi-vendor, multi-phase, and multi-disease cases, provides a diverse test bed to develop high-performance AI models for organ segmentation and tumor detection.

%%%%% fig:method
\begin{figure*}[t]
\centerline{\includegraphics[width=1.0\linewidth]{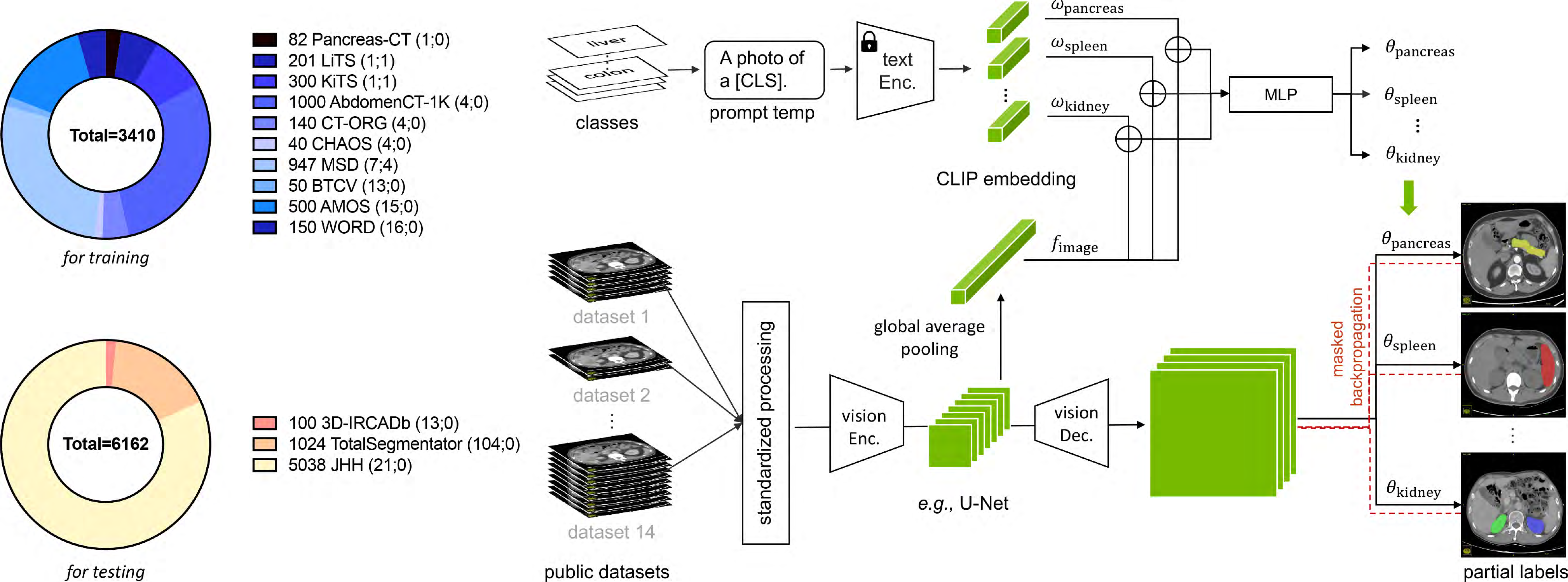}}
\caption{
\textbf{Overview.} We have developed a Universal Model from an assembly of 14 public datasets of 3,410 CT scans. In total, 25 organs and 6 types of tumors are partially labeled (detailed in~Appendix~\tableautorefname~\ref{tab:public_dataset}). To deal with partial labels, Universal Model consists of a text branch and a vision branch (\S\ref{sec:universal_model}). The official test set of MSD and BTCV are used to benchmark the performance of organ segmentation (\S\ref{sec:strong_challenge_ranking}) and tumor detection (\S\ref{sec:high_specificity}). 3D-IRCADb, TotalSegmentator and a large-scale private dataset, consisting of 5,038 CT scans with 21 annotated organs, are used for independent, external validation of model generalizability and transferability (\S\ref{sec:properties}). 
}
\label{fig:method}
\end{figure*}

\section{Related Work}
\label{sec:related_work}
\smallskip\noindent\textbf{\textit{Partial label problem.}}
Publicly available datasets for abdominal imaging focus on different organs and tumors~\cite{landman2017multiatlas,ma2021abdomenct,luo2021word,ji2022amos}, e.g., AbdomenCT-1K dataset for 4 organ segmentation~\cite{ma2021abdomenct}, WORD dataset for 16 organ segmentation~\cite{luo2021word} and TotalSegmentor dataset for 104 anatomical structure segmentation~\cite{wasserthal2022totalsegmentator}. 
The partial label problem occurs when training AI models on a combination of these datasets due to their inconsistent label taxonomy. 
To exploit the partial labels, several approaches have been investigated~\cite{zhou2019prior,fang2020multi,zhang2021dodnet,zhang2022merging}, aiming for a single model that can perform organ segmentation~\cite{liu2022universal,chen2021deep} and tumor detection~\cite{bai2022end,zlocha2019universal,yan2019mulan,liu2022improving,naga2022universal,yan2020universal,mattikalli2022universal}. These studies have the following limitations.
(1) Due to the small scale of the dataset assembly\footnote{{Zhou~\etal~\cite{zhou2019prior} assembled 150 CT scans from 4 datasets; Fang~\etal~\cite{fang2020multi} assembled 548 CT scans from 4 datasets; Zhang~\etal~\cite{zhang2021dodnet} assembled 1,155 CT scans from 7 datasets.}}, the potential of assembling datasets was not convincing. Their performance was similar to dataset-specific models and was not evaluated on the official benchmark.
(2) Due to the one-hot labels, the semantic relationship between organs and tumors was discarded.
\tableautorefname~\ref{tab:clip_embedding} reveals that the introduction of CLIP embedding is a salient factor to our proposed framework.

\smallskip\noindent\textbf{\textit{Organ segmentation and tumor detection.}}
Deep learning-based methods have been widely applied to organ segmentation and tumor detection. U-Net~\cite{ronneberger2015u} and its variants~\cite{zhou2019unet++,liang2021incorporating,oktay2018attention,isensee2021nnu} are one of the main streams and achieve some promising results. Recently, transformer based models \cite{chen2023towards,zhou2021nnformer,hatamizadeh2022unetr,tang2022self,chen2021transunet} are emerged, which can capture the global relationship between whole volume. These works are often specialized for single organ~\cite{ronneberger2015u,zhou2019unet++,isensee2021nnu,liang2021incorporating} or single task, i.e., organ segmentation~\cite{zhou2021nnformer,hatamizadeh2022unetr,tang2022self,chen2021transunet} or tumor detection~\cite{chen2023towards,yan2020learning,yan2018deeplesion}. Different from these work, Universal Model tackles both tasks within a single framework, using the introduced CLIP embedding to capture the semantic relationship between organs and tumors. Moreover, we demonstrate our work on publically available datasets, which is beneficial to reproducibility.

\smallskip\noindent\textbf{\textit{CLIP in medical imaging.}}
With the widespread success of large models in the field of language processing and understanding~\cite{devlin2018bert,brown2020language,singhal2022large,luddecke2022image}, large-scale pre-trained vision-language models (VLM), \eg, Conneau~\etal~\cite{conneau2019cross}, have recently been applied to multiple vision tasks~\cite{li2022language,rao2022denseclip,wang2022cris,chambon2022adapting,park2022per}, but rarely to the medical domain~\cite{eslami2021does,wang2022medclip}.
Qin~\etal~\cite{qin2022medical} suggested that VLM could be used for detection task in the medical domain with carefully designed medical prompts. 
Grounded in this findings, we are among the first to introduce CLIP embedding to voxel-level semantic understanding medical tasks, i.e., segmentation, in which we underline the importance of the semantic relationship between anatomical structures.

\smallskip\noindent\textbf{\textit{Medical universal models.}}
The field of medical image analysis has undergone a significant shift from training individual models for specific datasets towards developing a single (universal/foundation) model that can effectively handle diverse datasets, organs, tumors, tasks, and modalities.
After we first presented CLIP-Driven Universal Model in arXiv and released the code, the field has witnessed numerous pivotal contributions~\cite{yi2023towards,moor2023foundation,ulrich2023multitalent,huang2023stu,ye2023uniseg,butoi2023universeg,gao2023training,wang2023mis}, with many more endeavors underway to our knowledge~\cite{qu2023annotating,zhang2023continual}. Thereby, we are dedicated to reviewing the exceptional studies in the field by actively maintaining \href{https://github.com/ljwztc/CLIP-Driven-Universal-Model/blob/main/documents/awesome.md}{a GitHub page}. 

\section{Methodology}
\label{sec:method}

\subsection{Background}

\noindent\textbf{\textit{Problem definition.}} 
Let $M$ and $N$ be the total number of datasets to combine and data points in the combination of the datasets, respectively.
Given a dataset $\mathcal{D}=\{(\bm{X}_1,\bm{Y}_1), (\bm{X}_2,\bm{Y}_2), ..., (\bm{X}_N,\bm{Y}_N)\}$, there are a total of $K$ unique classes.
For $\forall n\in [1,N]$, if the presence of $\forall k\in [1,K]$ classes in $\bm{X}_i$ is annotated in $\bm{Y}_i$, $\mathcal{D}$ is a \textit{fully labeled} dataset; otherwise, $\mathcal{D}$ is a \textit{partially labeled} dataset.

\smallskip\noindent\textbf{\textit{Previous solutions.}} 
Two groups of solutions were proposed to address the partial label problem.
Given a data point $\bm{X}_n, n\in [1,N]$, the objective is to train a model $\mathcal{F}(\cdot)$ using the assembly dataset $\mathcal{D}_A = \{\mathcal{D}_1, \mathcal{D}_2, ..., \mathcal{D}_M\}$, and the model can predict all $K$ classes, if presented in $\bm{X}_n$.

\begin{itemize}
    \item \textit{Solution~\#1}~\cite{fang2020multi,shi2021marginal,yan2020learning,shi2021marginal,zhou2019prior,chen2019transfer,huang2020multi,tang2022self} aims to solve $\mathcal{F}_{\theta}(\bm{X}_n)=\bm{P}_n^k, n\in [1,N], k\in [1,K]$, where the prediction $\bm{P}_n$ is one-hot encoding with length $k$.

    \item \textit{Solution~\#2}~\cite{zhang2021dodnet,kang2021data,zhu2022assembling} aims to solve $\mathcal{F}_{\theta}(\bm{X}_n,\bm{w}_k)=\bm{P}_n, n\in [1,N], k\in [1,K]$,  where $\bm{w}_k$ is an one-hot vector to indicate which class to be predicted.
\end{itemize}

According to Zhang~\etal~\cite{zhang2021dodnet}, both solutions have similar segmentation performance, but \#2 is computationally more efficient. However, both solutions rely on one-hot labels, sharing two limitations. First, they ignore the semantic and anatomical relationship between organs and tumors. Second, they are inappropriate for segmenting various subtypes of tumors. To address these limitations, we modify $\bm{w}_k$ in Solution \#2 to CLIP embedding and introduce in-depth in the following sections.

%%%%%%% tab:clip_embedding
\begin{table}[t]
\caption
{\textbf{Label Encoding Ablation.} All three prompts can elicit knowledge from CLIP, achieving significant improvement over the conventional one-hot labels (DoDNet~\cite{zhang2021dodnet}) and BioBERT~\cite{yasunaga2022linkbert}. The average DSC score over validation part of Assembling Datasets is reported; per-class DSC found in Appendix \tableautorefname~\ref{tab:clip_ablation}.
}
\centering
\footnotesize
\begin{tabular}{p{0.18\linewidth}p{0.58\linewidth}P{0.08\linewidth}}
\hline
Embedding &  prompt & DSC \\
\shline
One-hot~\cite{zhang2021dodnet} & - & 70.42
\\
BioBERT~\cite{yasunaga2022linkbert} & {\scriptsize A computerized tomography of a [CLS].} & 71.55
\\
% CLIP V3 & {\scriptsize This computerized tomography has a [CLS].} & 66.76
% \\
CLIP V1 & {\scriptsize A photo of a [CLS].} & 73.49 \\
CLIP V2 & {\scriptsize There is [CLS] in this computerized tomography.} & 75.66
\\
CLIP V3 & {\scriptsize A computerized tomography of a [CLS].} & \textbf{76.11}
\\
\hline
\end{tabular}
\label{tab:clip_embedding}
\end{table}

\subsection{CLIP-Driven Universal Model}
\label{sec:universal_model}

The overall framework of CLIP-Driven Universal Model (see~\figureautorefname~\ref{fig:method}) has a text branch and a vision branch. The text branch first generates the CLIP embedding for each organ and tumor using an appropriate medical prompting (\tableautorefname~\ref{tab:clip_embedding}), and then the vision branch takes both CT scans and CLIP embedding to predict the segmentation mask\footnote{Our framework design is conceptually similar to \textit{Segment Anything Model (SAM)}~\cite{kirillov2023segment}, which is a concurrent study of ours in computer vision. By leveraging CLIP embedding as a prompt within our Universal Model, we are able to generate highly accurate masks for organs and tumors of interest, as opposed to producing masks for arbitrary objects.}.

\smallskip\noindent\textbf{\textit{Text branch.}}
Let $\bm{w}_k$ be the CLIP embedding of the $k$-th class, produced by the pre-trained text encoder in CLIP and a medical prompt (\eg, ``a computerized tomography of a [CLS]'', where [CLS] is a concrete class name).
We first concatenate the CLIP embedding ($\bm{w}_k$) and the global image feature ($\bm{f}$) and then input it to a multi-layer perceptron (MLP), namely \textit{text-based controller}~\cite{tian2020conditional},  to generate parameters ($\boldsymbol{\theta}_{k}$), \ie,
$\boldsymbol{\theta}_{k}=\text{MLP}(\bm{w}_k \oplus \bm{f})$,
where $\oplus$ is the concatenation.
Although CLIP embedding significantly outperforms one-hot labels~\cite{zhang2021dodnet}, we mark that the choice of medical prompt template is critical.  \tableautorefname~\ref{tab:clip_embedding} presents the effectiveness of three prompt templates. 
Moreover, the introduction of CLIP embedding addresses the label orthogonality problem by exploiting semantic relationships among organs and tumors (illustrated in~\figureautorefname~\ref{fig:teaser}).
%The CLIP embedding also allows us to extend the Universal Model to novel classes (\eg, organs, tumors, bones, and other anatomical structures) because the length of CLIP embedding is fixed and the incremental learning of new classes will not affect other classes.

\smallskip\noindent\textbf{\textit{Vision branch.}} We pre-process CT scans using isotropic spacing and uniformed intensity scale to reduce the domain gap among various datasets\footnote{A standardized and normalized CT pre-processing is important when combining multiple datasets.
Substantial differences in CT scans can occur in image quality and technical display, originating from different acquisition parameters, reconstruction kernels, contrast enhancements, intensity variation, and so on~\cite{orbes2019multi,yan2020mri,guo2021multi}.
\iffalse {\jlred In our assembly of 14 public datasets, in-plane image resolution of the axial scans varies from 0.56 mm to 1.0 mm and the slice thickness varies from 0.45 mm to 6.0 mm. The number of slices per volume ranges from 42 to 128.}\fi}. The standardized and normalized CT scans are then processed by the vision encoder. Let $\bm{F}$ be the image features extracted by the vision encoder. 
To process $\bm{F}$, we use three sequential convolutional layers with $1\times1\times1$ kernels, namely \textit{text-driven segmentor}. 
The first two layers have 8 channels, and the last one has 1 channel, corresponding to the class of [CLS]$_k$. 
% Thus, a total of 153 parameters are required from the text-based controller output. 
The prediction for the class [CLS]$_k$ is computed as
$\bm{P}_{k}=\text{Sigmoid}\left(\left(\left(\bm{F} * \boldsymbol{\theta}_{k_1}\right) * \boldsymbol{\theta}_{k_2}\right) * \boldsymbol{\theta}_{k_3}\right)$,
where $\boldsymbol{\theta}_{k} = \{ \boldsymbol{\theta}_{k_1}, \boldsymbol{\theta}_{k_2}, \boldsymbol{\theta}_{k_3} \}$ are computed in the text branch, and $*$ represents the convolution. For each class [CLS]$_k$, we generate the prediction $\bm{P}_{k} \in \mathbb{R}^{1 \times D \times W \times H}$ representing the foreground of each class in \textit{one vs. all} manner (\ie, Sigmoid instead of Softmax).
%When extending to upcoming organs, it is easily enlarge the dictionary with new organ name and keep the network architecture to finetune the whole framework.

\smallskip\noindent\textbf{\textit{Masked back-propagation.}} To address the label inconsistency problem, we proposed the masked back-propagation technique. The BCE loss function is utilized for supervision. We masked the loss terms of these classes that are not contained in $\bm{Y}$ and only back-propagate the accurate supervision to update the whole framework. 
The masked back-propagation addresses the label inconsistency in the partial label problem. Specifically, partially labeled datasets annotate some other organs as background, leading to the disability of existing training schemes (Solution \#1).

%%%%%%% tab:msd_test
\begin{table*}[t]
\caption{\textbf{Leaderboard performance on MSD.} The results are evaluated in the server on the MSD competition test dataset. All Dice and NSD metrics are obtained from the \href{https://decathlon-10.grand-challenge.org/evaluation/challenge/leaderboard/}{MSD public leaderboard}. The results of MRI-related tasks were generated by Swin UNETR~\cite{tang2022self}.}
 \centering
\footnotesize
\begin{tabular}{p{0.14\linewidth}|P{0.045\linewidth}P{0.045\linewidth}P{0.045\linewidth}|P{0.045\linewidth}P{0.045\linewidth}P{0.045\linewidth}|P{0.045\linewidth}P{0.045\linewidth}P{0.045\linewidth}|P{0.045\linewidth}P{0.045\linewidth}P{0.045\linewidth}}
\hline
 & \multicolumn{6}{c|}{Task03 Liver} & \multicolumn{6}{c}{Task07 Pancreas}  \\
\cline{2-13}
Method & Dice1 & Dice2 &Avg. & NSD1  & NSD2 &Avg. & Dice1  & Dice2  & Avg.  & NSD1 & NSD2 & Avg. \\ 
\shline
Kim~\etal~\cite{kim2019scalable}  & 94.25 & 72.96   & 83.61  & 96.76 & 88.58 & 92.67 & 80.61    & 51.75  & 66.18 & 95.83   & 73.09  & 84.46 \\
Trans VW~\cite{haghighi2021transferable}  & 95.18 & 76.90 & 86.04   & 97.86  & {92.03} & 94.95 & 81.42    & 51.08  & 66.25 & 96.07 & 70.13   & 83.10 \\
C2FNAS\cite{yu2020c2fnas}  & 94.98 & 72.89   & 83.94  & 98.38 & 89.15 & 93.77  & 80.76    & 54.41  & 67.59 & 96.16   & 75.58  & 85.87 \\
Models Gen.~\cite{zhou2021models}   & 95.72 & {77.50}   & {86.61}  & 98.48 & 91.92 & {95.20} & 81.36    & 50.36  & 65.86 & 96.16   & 70.02  & 83.09 \\
nnUNet~\cite{isensee2021nnu}  & \textbf{95.75} & 75.97   & 85.86  & 98.55 & 90.65 & 94.60 &  81.64    & 52.78  & 67.21 & 96.14   & 71.47  & 83.81 \\
DiNTS~\cite{he2021dints}   & 95.35 & 74.62   & 84.99  & \textbf{98.69} & 91.02 & 94.86  & 81.02    & 55.35  & 68.19 & 96.26   & 75.90  & 86.08 \\
Swin UNETR~\cite{tang2022self}   & 95.35 & 75.68 & 85.52 & 98.34 & 91.59 & 94.97 &  {81.85} &{58.21} & {70.71} & {96.57} &{79.10} & {87.84} \\
\hline
Universal Model & 95.42 & \textbf{79.35} & \textbf{87.39} & 98.18 & \textbf{93.42} & \textbf{95.80} & \textbf{82.84} & \textbf{62.33} & \textbf{72.59} & \textbf{96.65} & \textbf{82.86} & \textbf{89.76} 
\\

\hline

\end{tabular}
\vspace{0.3 em}\\
\begin{tabular}{p{0.14\linewidth}|P{0.045\linewidth}P{0.045\linewidth}P{0.045\linewidth}|P{0.045\linewidth}P{0.045\linewidth}P{0.045\linewidth}|P{0.045\linewidth}P{0.045\linewidth}|P{0.045\linewidth}P{0.045\linewidth}|P{0.045\linewidth}P{0.045\linewidth}}
\hline
& \multicolumn{6}{c|}{Task08 Hepatic Vessel} & \multicolumn{2}{c|}{Task06 Lung} & \multicolumn{2}{c|}{Task09 Spleen} & \multicolumn{2}{c}{Task10 Colon} \\
\cline{2-13}
Method & Dice1  & Dice2  & Avg.  & NSD1 & NSD2 & Avg. & Dice1 & NSD1 & Dice1 & NSD1 & Dice1 & NSD1\\ 
\shline
Kim~\etal~\cite{kim2019scalable} & 62.34 & 68.63   & 65.49  & 83.22 & 78.43 & 80.83 & 63.10 & 62.51 & 91.92    & 94.83 & 49.32    & 62.21\\
Trans VW~\cite{haghighi2021transferable}  & 65.80 & 71.44 & 68.62   & 84.01  & 80.15 & 82.08 &74.54    & 76.22 & 97.35    & 99.87 & 51.47    & 60.53\\
C2FNAS\cite{yu2020c2fnas}  &  64.30 & 71.00   & 67.65  & 83.78 & 80.66 & 82.22  &70.44    & 72.22 & 96.28    & 97.66 & 58.90    & {72.56}\\
Models Gen.~\cite{zhou2021models} & 65.80 & 71.44   & 68.62  & 84.01 & 80.15 & 82.08 & 74.54    & 76.22  & 97.35    & 99.87 & 51.47    & 60.53\\
nnUNet~\cite{isensee2021nnu}  & {66.46} & 71.78   & {69.12}  & 84.43 & 80.72 & 82.58 &73.97    & 76.02  & \textbf{97.43}    & \textbf{99.89} & 58.33    & 68.43\\
DiNTS~\cite{he2021dints}  &  64.50 & 71.76   & 68.13  & 83.98 & 81.03 & 82.51  &74.75    & 77.02 & 96.98    & 99.83 & 59.21    & 70.34\\
Swin UNETR~\cite{tang2022self} & 65.69 & {72.20} &68.95 & {84.83} & {81.62} & {83.23}  &{76.60} &{77.40} & 96.99 & 99.84 & {59.45} & 70.89\\
\hline
Universal Model & \textbf{67.15} & \textbf{75.86} & \textbf{71.51} & \textbf{84.84} & \textbf{85.23} & \textbf{85.04} & \textbf{80.01} & \textbf{81.25} & 97.27 & 99.87 & \textbf{63.14} & \textbf{75.15}
\\
\hline
\end{tabular}%
\label{tab:msd_test}
\end{table*}

%%%%%%% fig:msd_benchmark
\begin{figure*}[h!]
\centering
\footnotesize
    \includegraphics[width=0.95\linewidth]{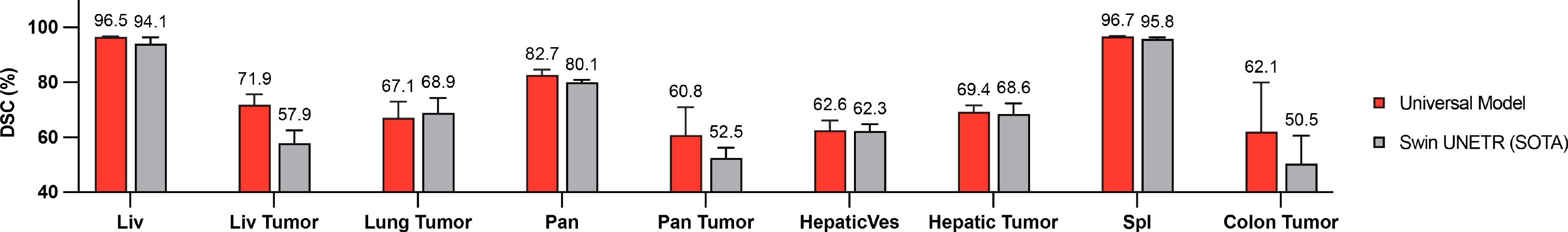}
\caption{
    \textbf{Benchmark on MSD validation dataset.} We compare Universal Model with Swin UNETR~\cite{tang2022self} (previously ranked first on the MSD leaderboard) on 5-fold cross-validation of the MSD dataset. Universal Model achieves overall better segmentation performance and offers \textit{substantial} improvement in the tasks of segmenting liver tumors (+14\%), pancreatic tumors (+8\%), and colon tumors (+11\%).
}
% \vspace{-0.2cm}
\label{fig:msd_benchmark}
\end{figure*}

%%%%% tab:btcv_benchmark
\begin{table*}[t]
\caption
{\textbf{5-fold cross-validation results on BTCV.} For a fair comparison, we did not use model ensemble during the evaluation. All experiments are under the same data splits, computing resources, and testing conditions. Universal Model achieves the overall best performance, yielding at least +3.9\% DSC improvement over the state-of-the-art method.}
\centering
\footnotesize
\begin{tabular}{p{0.12\linewidth}P{0.04\linewidth}P{0.04\linewidth}P{0.04\linewidth}P{0.04\linewidth}P{0.04\linewidth}P{0.04\linewidth}P{0.04\linewidth}P{0.04\linewidth}P{0.04\linewidth}P{0.04\linewidth}P{0.04\linewidth}P{0.04\linewidth}|P{0.06\linewidth}}
\hline
Methods & Spl  
& RKid & LKid 
& Gall  & Eso 
& Liv & Sto 
& Aor & IVC 
& Veins   & Pan 
& AG & Avg. \\ \shline
% % ASPP~\cite{deeplabv3plus2018}                   
% & 94.19 & 91.24                        
% & 88.02 & 63.58                        
% & 72.64 & 93.61                       
% & 80.05 & 86.20   
% & 80.11 & 71.49
% & 74.29 & 64.58
% & 78.81
% \\ 
% PaNN~\cite{zhou2019prior}     
% & 94.04 & 90.21                      
% & 88.58 & 64.96                       
% & 73.20 & 93.50                     
% & 80.87 & 87.26  
% & 80.32 & 70.59
% & 75.28 & 64.92
% & 79.13
% \\
% TransUNet~\cite{chen2021transunet}    
% & 94.10 & 90.22                        
% & 88.84 & 65.49                    
% & 73.19 & 93.24                        
% & 80.85 & 87.47   
% & 80.48 & 71.47
% & 74.26 & 64.76
% & 79.16
% \\ 
% CoTr*~\cite{xie2021cotr}     
% & 95.60 & 89.22                      
% & 88.58 & 67.54                        
% & 74.95 & 95.97                        
% & 81.95 & 88.58   
% & 81.20 & 72.83
% & 76.69 & 65.81
% & 80.36
% \\ 
% CoTr~\cite{xie2021cotr}     
% & 95.51 & 88.03                 
% & 89.19 & 68.49                        
% & 75.83 & 95.93                       
% & 81.84 & 89.01 
% & 82.32 & 73.39
% & 75.12 & 65.78
% & 80.48
% \\
RandPatch~\cite{tang2021high}     
& 95.82 & 88.52                        
& 90.14 & 68.31                        
& 75.01 & 96.48                        
& 82.93 & 88.96   
& 82.49 & 73.54
& 75.48 & 66.09
& 80.76
\\ 
TransBTS~\cite{isensee2021nnu}     
& 94.59 & 89.23                     
& 90.47 & 68.50                       
& 75.59 & 96.14                       
& 83.72 & 88.85  
& 82.28 & 74.25
& 75.12 & 66.74 
& 80.94
\\ 
nnFormer~\cite{zhou2021nnformer}     
& 94.51 & 88.49                     
& 93.39 & 65.51                        
& 74.49 & 96.10                      
& 83.83 & 88.91
& 80.58 & 75.94
& 77.71 & {68.19}
& 81.22
\\
UNETR~\cite{hatamizadeh2022unetr}
& 94.91 & 92.10                        
& 93.12 & 76.98                        
& 74.01 & 96.17                        
& 79.98 & 89.74   
& 81.20 & 75.05
& 80.12 & 62.60
& 81.43
\\
nnU-Net~\cite{isensee2021nnu}     
& \textbf{95.92} & 88.28                        
& 92.62 & 66.58                        
& 75.71 & 96.49                        
& 86.05 & 88.33   
& 82.72 & \textbf{78.31}
& 79.17 & 67.99
& 82.01
\\
Swin UNETR~\cite{tang2022self}    
& 95.44 & {93.38}                 
& 93.40 & 77.12                    
& 74.14 & 96.39                        
& 80.12 & 90.02   
& 82.93 & 75.08
& 81.02 & 64.98
& 82.06
\\
\hline
Universal Model & 95.82 & \textbf{94.28} & \textbf{94.11} & \textbf{79.52} & \textbf{76.55} & \textbf{97.05} & \textbf{92.59} & \textbf{91.63} & \textbf{86.00} & 77.54 & \textbf{83.17} & \textbf{70.52} & \textbf{86.13}
\\ \hline
\end{tabular}%
\label{tab:btcv_benchmark}
\end{table*}

% \subsection{Assembling Public Datasets}
% \label{sec::assembly_of_public_dataset}

% To fully explore the potential of the proposed Universal Model in multi-organ segmentation and tumor detection, we have thus far assembled a total of 3,410 CT scans from 14 public datasets (many more can be included when datasets are available).
% However, it is non-trivial to assemble partially labeled datasets. Apart from the label inconsistency and label orthogonality that we have addressed, two other open challenges are discussed in Appendix~\S\ref{sec:open_challenges_review}.

%%%%% fig:pseudo_truth_evaluation
\begin{figure}[t]
\centerline{\includegraphics[width=\linewidth]{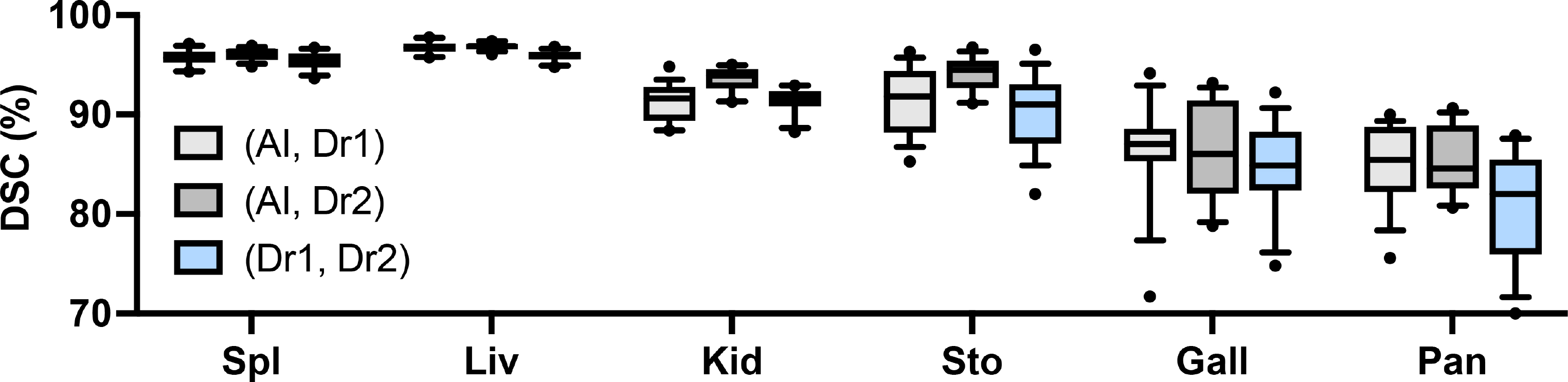}}
\caption{
\textbf{Intra-observer variability.} We obtain similar performance between pseudo labels generated by the Universal Model (AI) and annotations performed by two human experts (Dr1,2) on 6 organs. Spleen (Spl), liver (Liv), kidneys (Kid), stomach (Sto), gallbladder (Gall), and pancreas (Pan) can be annotated by AI with a similar intra-observer variability to humans. Examples of pseudo labels and human annotations are provided in Appendix \figureautorefname~\ref{fig:pseudo_truth_visualization}.
}
\label{fig:pseudo_truth_evaluation}
% \vspace{-0.2cm}
\end{figure}

%%%%% fig:qualitative_visualization
\begin{figure*}[t]
\centerline{\includegraphics[width=1.0\linewidth]{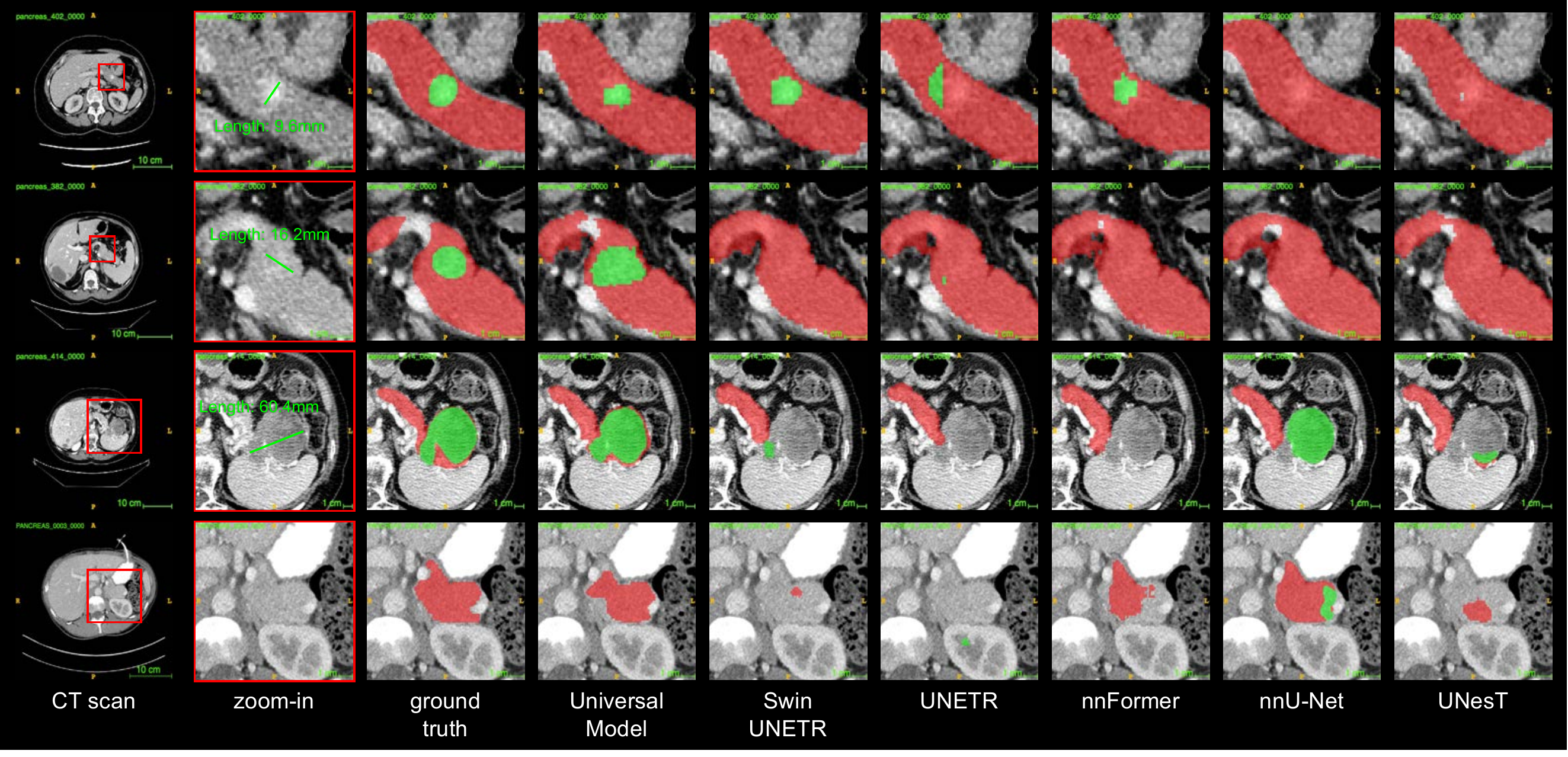}}
\caption{
\textbf{Pancreatic tumor detection.} Qualitative visualizations of the proposed Universal Model and five competitive baseline methods. We review the detection results of tumors from smaller to larger sizes (Rows 1--3). When it comes to a CT scan without tumor from other hospitals, the Universal Model generalize well in organ segmentation and does not generate many false positives of tumors (Row 4; \S\ref{sec:high_specificity}). The visualization of tumor detection in other organs (\eg, liver tumors and kidney tumors) can be found in Appendix~Figures~\ref{fig:qualitative_visualization_liver_tumor}--\ref{fig:qualitative_visualization_kidney_tumor}.
}
\label{fig:qualitative_visualization}
% \vspace{-0.2cm}
\end{figure*}

%%%%% tab:high_specificity
\begin{table*}[t]
\caption
{\textbf{Tumor detection performance.} The CT scans in LiTS~\cite{bilic2019liver}, KiTS~\cite{heller2019kits19}, and MSD Pancreas~\cite{antonelli2021medical} contain tumors in the liver, kidney, and pancreas, respectively. These scans are used to compute the sensitivity (Sen.) of tumor detection. To perform an alternative check of specificity (Spec.), we use CHAOS~\cite{valindria2018multi} and Pancreas-CT~\cite{roth2015deeporgan}. It has been confirmed that CHAOS has no liver or kidney tumor, and Pancreas-CT has no pancreatic tumor in the CT scans. The harmonic mean (Harm.) is calculated to indicate the balance between sensitivity and specificity. Universal Model achieves high harmonic mean, which is clinically important because it reveals that Universal Model can accurately identify tumor cases while reduce false positives.}
\centering
\footnotesize
\begin{tabular}{p{0.18\linewidth}|P{0.07\linewidth}P{0.07\linewidth}P{0.05\linewidth}|P{0.07\linewidth}P{0.07\linewidth}P{0.05\linewidth}|P{0.07\linewidth}P{0.07\linewidth}P{0.05\linewidth}}
\hline
\multirow{2}{*}{Methods} & \multicolumn{3}{c|}{Liver Tumor} & \multicolumn{3}{c|}{Kidney Tumor} &\multicolumn{3}{c}{Pancreatic Tumor} \\
& Sen. & Spec. & Harm.
& Sen. & Spec. & Harm.
& Sen. & Spec. & Harm.
\\ \shline
nnU-Net~\cite{isensee2021nnu} & \textbf{94.44} & 75.00 & 83.60 & 96.88 & 85.00 & 90.55 & 95.18 & 88.75 & 91.85
\\
UNet++~\cite{zhou2019unet++}& \textbf{94.44} & 80.00 & 86.62 &N/A&N/A&N/A&N/A & N/A&N/A
%\\
%Models Gen.~\cite{zhou2021models}
% \\
% TransUNet~\cite{chen2021transunet} &N/A&N/A&N/A&N/A&N/A&90.00
\\
UNETR~\cite{hatamizadeh2022unetr} & 86.11 & \textbf{95.00} & 90.34 & 93.75 & \textbf{95.00} & \textbf{94.37} & 90.36 & 81.25 & 85.56
\\
Swin UNETR~\cite{tang2022self} & 91.67 & 85.00 & 88.21 & \textbf{97.91} & 70.00 & 81.63 & \textbf{97.59} & 87.50 & 92.26
\\
\hline
Universal Model & 88.89 & \textbf{95.00} & \textbf{91.84} & 91.67 & \textbf{95.00} & 93.31 & 93.98 & \textbf{91.25} & \textbf{92.59}
\\ \hline
\end{tabular}%
\label{tab:high_specificity}
\end{table*}

%%%%% fig:pseudo_tsne
\begin{figure}[t]
\centerline{\includegraphics[width=1.0\linewidth]{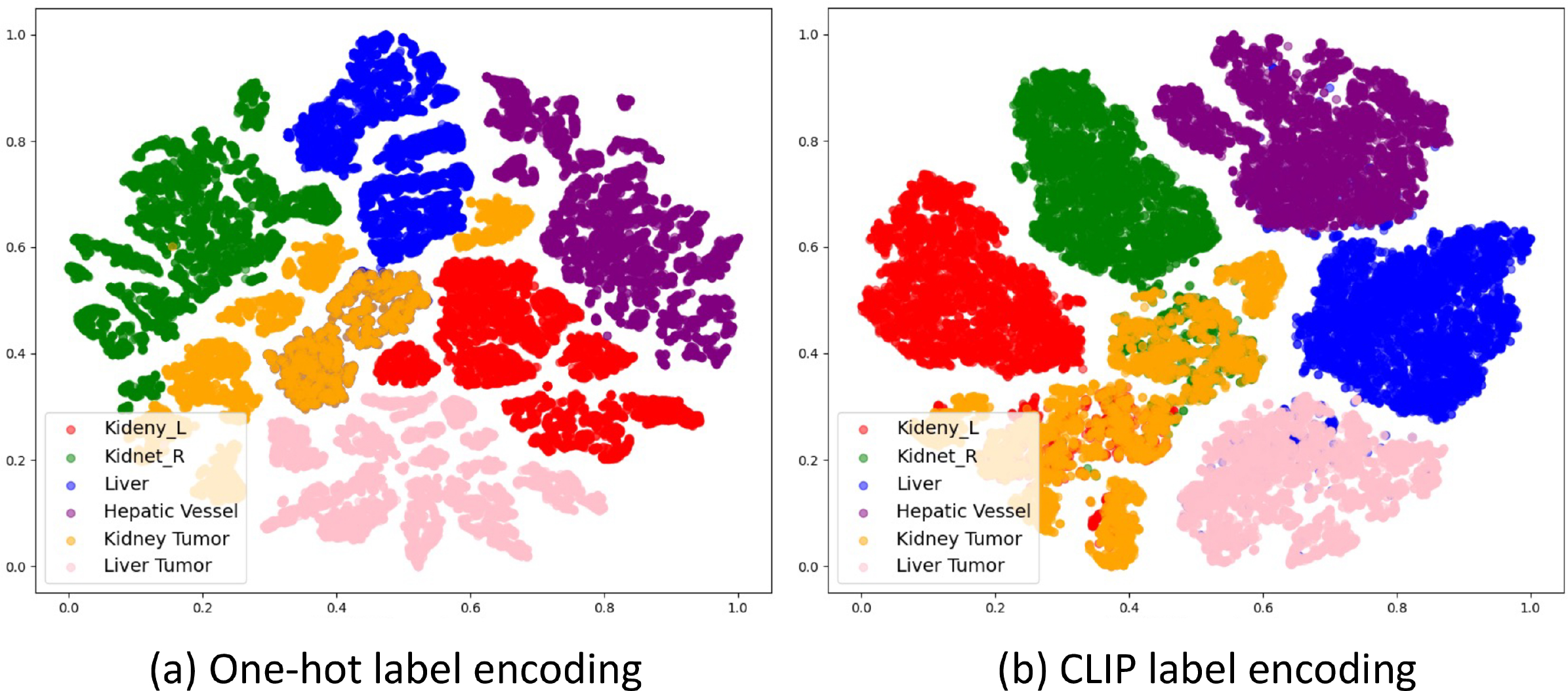}}
\caption{
\textbf{t-SNE visualization of embedding space.} We compare the decoder embedding space of (a) One-hot label encoding and (b) CLIP label encoding with six categories, i.e., liver, liver tumor, right kidney, left kidney, kidney tumor and hepatic vessel, which is the same as in \figureautorefname~\ref{fig:teaser}. CLIP label encoding achieves a better feature cluster and shows anatomically structured semantics.
Visualization of embedding space for all categories is provided in Appendix \figureautorefname~\ref{fig:all_tsne}.
}
\label{fig:tsne}
\end{figure}

\iffalse
%%%%%%% tab:expansibility
\begin{table}[t]
\caption
{\textit{\jlred Expansibility:} Flexible Backbone.
}

\footnotesize
\begin{tabular}{p{0.30\linewidth}|P{0.15\linewidth}|P{0.15\linewidth}|P{0.15\linewidth}}
\hline
Backbone &  Organ & Tumor & Average \\
\shline
SwinUNETR~\cite{tang2022self} &79.56 & 63.82 & 76.11
\\
UNet~\cite{ronneberger2015u} & 79.66  & 66.28 & 76.73\\
%UNet++~\cite{zhou2019unet++} &  & 
%\\
%Dints &  & 
\hline
\end{tabular}
\label{tab:expansibility}
\end{table}
\fi

%%%%% fig:computational_efficiency
\begin{figure}[h]
\centerline{\includegraphics[width=0.96\linewidth]{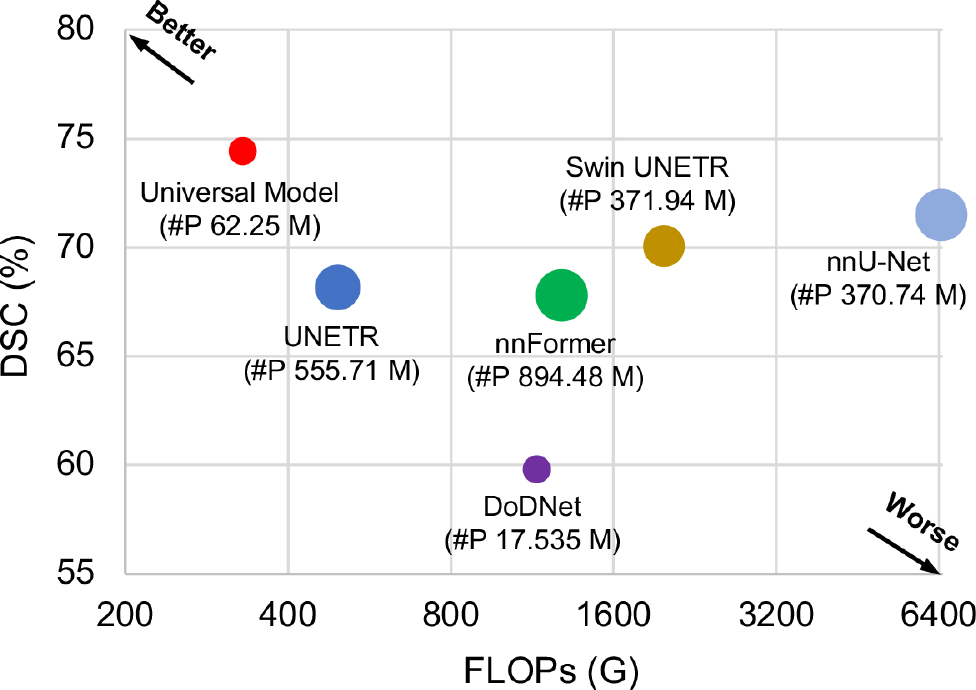}}
\caption{
\textbf{\textit{Efficiency:} FLOPs vs. DSC.} We plot the average DSC score on the 6 MSD tasks against the FLOPs (Floating-point operations per second). The FLOPs is computed based on input with spatial size $96\times96 \times 96$. The size of each circle indicates the number of parameters (`\#P'). In the inference, Universal Model is faster than nnU-Net (2nd best in performance) and Swin UNETR (3rd best) by 19$\times$ and 6$\times$ measured by FLOPs, respectively.
}
\label{fig:computational_efficiency}
\end{figure}

%%%%% fig:pseudo_tsne
\begin{figure}[t]
\centerline{\includegraphics[width=1.0\linewidth]{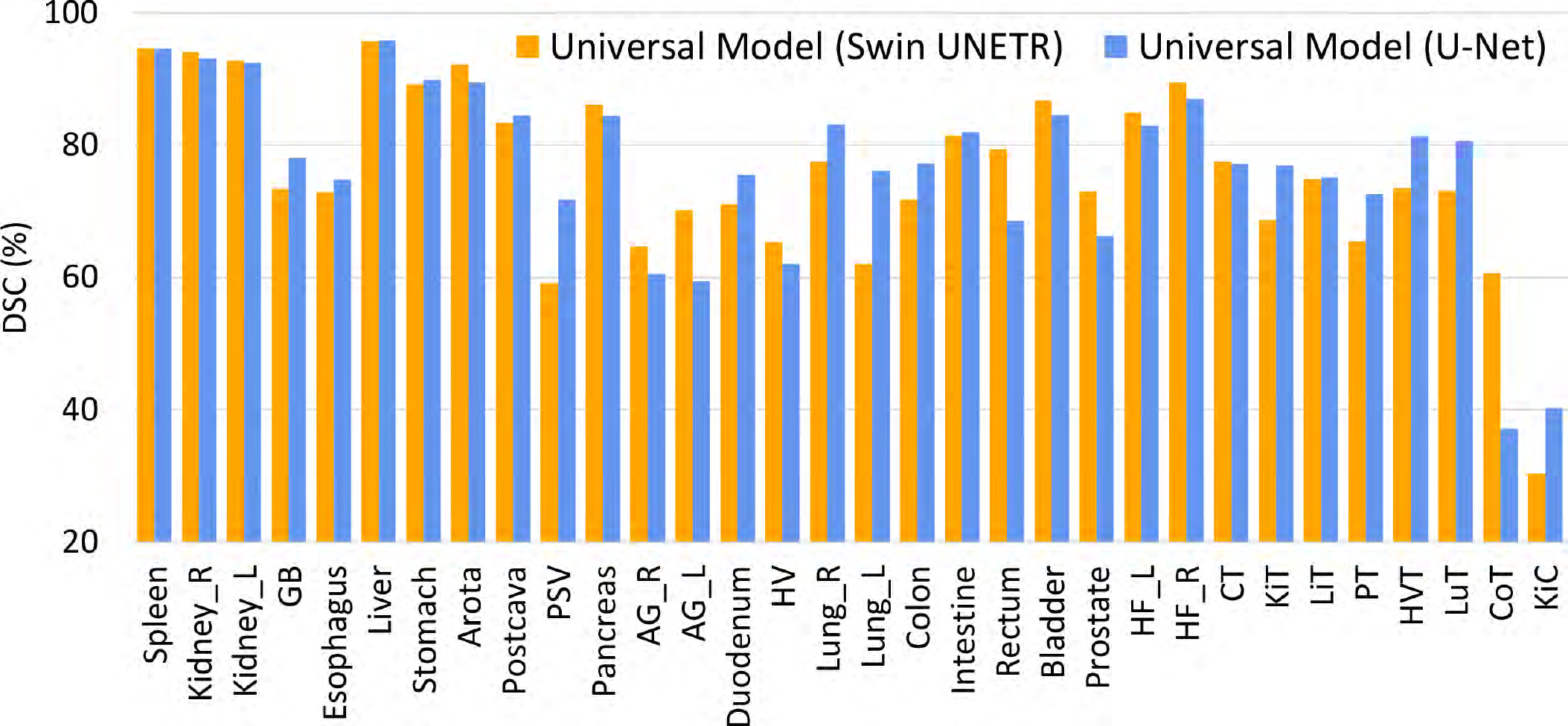}}
\caption{
\textbf{\textit{Expansibility:} flexible backbones.} Universal Model can be expanded to CNN-based (\eg, U-Net~\cite{ronneberger2015u}) and Transformer-based (\eg, Swin UNETR~\cite{tang2022self}) backbone. For the abbreviation of some organs, please refer to Appendix \tableautorefname~\ref{tab:clip_ablation}. Both backbones achieve comparable results.
}
\label{fig:expansibility}
% \vspace{-0.2cm}
\end{figure}

\section{Experiments \& Results}
\label{sec:result}

\noindent\textbf{\textit{Datasets and evaluation.}} 14 public datasets of 3,410 CT scans in total are assembled for training. Other two public and one private datasets are used for testing. Dataset details and pre-processing are in Appendix~\S\ref{sec:assembly_datasets}. Dice Similarity Coefficient (DSC) and Normalized Surface Distance (NSD) are evaluated for organ/tumor segmentation; Sensitivity and Specificity are for tumor detection. 

\smallskip\noindent\textbf{\textit{Implementation details.}} The Universal Model is trained using the AdamW optimizer with a warm-up cosine scheduler of 50 epochs. The segmentation experiments use batch-size of 6 per GPU with a patch size of $96\times96\times96$. Default initial learning rate of $4e^{-4}$, momentum of 0.9 and decay of $1e^{-5}$ on multi-GPU (4) with DDP. The framework is implemented in MONAI 0.9.0\footnote{\href{https://monai.io/}{https://monai.io/}}. The five-fold cross validation strategy is performed. We select the best model in each fold by evaluating the validation best metrics. Models are trained on eight NVIDIA RTX A5000 cards.

\subsection{Organ Segmentation on MSD and BTCV}
\label{sec:strong_challenge_ranking}
We offer the top \#1 solution in both Medical Segmentation Decathlon (MSD)\footnote{\href{https://decathlon-10.grand-challenge.org/evaluation/challenge/leaderboard/}{decathlon-10.grand-challenge.org/evaluation/challenge/leaderboard/}} and Beyond The Cranial Vault (BTCV), surpassing the runners-up by a considerable margin. It's noted that universal model provides six CT tasks solution and the results of other four MRI tasks are predicted by nnUnet \cite{isensee2021nnu}.
\tableautorefname~\ref{tab:msd_test} and \figureautorefname~\ref{fig:msd_benchmark} present detailed comparison on the official test set and 5-hold cross validation on MSD, respectively.
\tableautorefname~\ref{tab:btcv_benchmark} compares Universal Model with other methods in the validation set of BTCV, offering at least 3.5\% improvements over the second best.

Manual annotations have inter-rater and intra-rater variance~\cite{ji2021learning}, particularly in segmentation tasks, because some of the organs' boundaries are blurry and ambiguous.
We assess the quality of pseudo labels predicted by Universal Model and manual annotation performed by human experts.
17 CT scans in BTCV have been annotated by two independent groups of radiologists from different institutes (not test server labels).
As a result, each CT scan is associated with AI prediction, and two human annotations (Dr1 and Dr2).
\figureautorefname~\ref{fig:pseudo_truth_evaluation} presents their mutual DSC scores, \ie, AI$\leftrightarrow$Dr1, AI$\leftrightarrow$Dr2, and Dr1$\leftrightarrow$Dr2.
We find the DSC between AI and humans is slightly larger than the DSC between humans in segmenting 6 types of organs (\ie, spleen, liver, kidney, stomach, and pancreas).
With this high-quality AI prediction, we assemble a large dataset of 3,410 CT scans from a diverse set of hospitals (\figureautorefname~\ref{fig:method} and generate pseudo labels for 25 organs and 6 tumors\footnote{The quality of 19 other organs and 6 tumors has not been compared with human annotations because there is no publicly available CT scans that have been annotated by multiple independent groups on these objects.}. Pseudo-label refinement has been performed for a few CT scans where AI's prediction is uncertain.
This fully annotated dataset will be released (examples in Appendix~\figureautorefname~\ref{fig:multiorgan_multitumor}).
Now that these 6 organs can be segmented by AI with a similar variance to human experts, we encourage the research community to concentrate on creating annotations for harder organs and tumors.

\subsection{Tumor Detection on Five Datasets}
\label{sec:high_specificity}

\figureautorefname~\ref{fig:msd_benchmark} demonstrates that Universal Model surpasses Swin UNETR by a large margin in segmenting liver, pancreatic, and colon tumors, leading to 14\%, 8\%, and 12\% improvement in DSC scores, respectively.
However, DSC scores cannot faithfully reveal the tumor detection performance because, by default, they are only calculated on abnormal CT scans (with tumors)~\cite{isensee2021nnu}. The AI might generate numerous false positives when encountering normal CT scans (that have no tumor)~\cite{shen2021artificial}.
Therefore, we also evaluate patient-level Sensitivity and Specificity for detecting the three types of tumors, and the harmonic mean of sensitivity and specificity is reported to indicate the balance between two abilities.
To obtain normal CT scans, we adopt the CHAOS and Pancreas-CT datasets because these two datasets provide pathological verification that no tumors are present~\cite{valindria2018multi,roth2015deeporgan}.
\tableautorefname~\ref{tab:high_specificity} show that Universal Model achieves harmonic mean of 91.84\%, 93.31\% and 92.59\% for three tumors, indicating the ability to accurately identify tumor cases while reducing false positives and achieving a competitive balance.
Moreover, Rows 1--3 in~\figureautorefname~\ref{fig:qualitative_visualization} depict the prediction of small/medium/large pancreatic tumors; Row 4 shows that Universal Model can precisely segment the pancreas and reduce the number of false positives on normal CT scans.
Compared with dataset-specific models, the smaller number of false positives predicted by our Universal Model underlines the necessity of assembling diverse datasets, benefiting from not only sufficient positive examples for training but also a larger number of negative examples as a control. 

\subsection{Effectiveness of CLIP Embedding}
We further show the t-SNE visualization of embedding space for both one-hot encoding and CLIP encoding in \figureautorefname~\ref{fig:tsne}. We can see that the decoder embedding of CLIP encoding shows better feature clustering and anatomical structure. For example, right kidney and left kidney features are closer in embedding space for universal model, which is highly matched with cosine similarity between CLIP embeddings as shown in \figureautorefname~\ref{fig:teaser}. This validates that the CLIP-based encoding can facilitate the model to capture the anatomical relationship and to learn a structured feature embedding. 
Furthermore, we conduct ablation study with various embedding to replace the CLIP embedding including BioLinkBERT embedding\footnote{LinkBERT is a transformer encoder model pretrained on a large corpus of documents, which has capabilities for understanding medical text.} ~\cite{yasunaga2022linkbert}, and the results are shown in Appendix \tableautorefname~\ref{tab:clip_ablation}. We can see that the CLIP-based embedding can significantly improve the performance comparing with conventional one-hot labels (DoDNet \cite{zhang2021dodnet}) and text-only pre-trained embedding (BioLinkBERT \cite{yasunaga2022linkbert}).

%%%%% tab:generalizability
\begin{table*}[h]
\caption{\textbf{\textit{Generalizability:} Results on external datasets.} We evaluate Universal Model and eight models on data from two external sources without additional fine-tuning or domain adaptation. {mDSC* is the average dice score of the first seven organs.} Compared with dataset-specific models, our Universal Model performs more robustly to CT scans taken from a variety of scanners, protocols, and institutes.}
\centering
\footnotesize
\begin{tabular}{p{0.15\linewidth}|P{0.05\linewidth}P{0.05\linewidth}P{0.05\linewidth}P{0.05\linewidth}P{0.05\linewidth}P{0.05\linewidth}P{0.05\linewidth}P{0.05\linewidth}P{0.05\linewidth}P{0.05\linewidth}P{0.05\linewidth}}
\hline
\textbf{\textit{3D-IRCADb}} & spleen & kidneyR & kidneyL
& gallbladder & liver & stomach 
& pancreas & lungR & lungL & mDSC* & mDSC
\\ \shline
% nnU-Net~\cite{isensee2021nnu} & 93.98 & 94.15 & 94.75 & 64.14 & 95.91 & 89.05 & 80.43 & N/A & N/A & N/A & 87.48
% \\
SegResNet~\cite{siddiquee2021redundancy} & 94.08 & 80.01 & 91.60 & 69.59 & 95.62 & \textbf{89.53} & 79.19 & N/A & N/A & 85.66 & N/A
\\
nnFormer~\cite{zhou2021nnformer} & 93.75 & 88.20 & 90.11 & 62.22 & 94.93 & 87.93 & 78.90  & N/A & N/A & 85.14 & N/A
\\
UNesT~\cite{yu2022unest} & 94.02 & 84.90 & \textbf{94.95} & 68.58 & 95.10 & 89.28 & 79.94 & N/A & N/A & 86.68 & N/A
\\
TransBTS~\cite{wang2021transbts} & 91.33 & 76.22 & 88.87 & 62.50 & 94.42 & 85.87 & 63.90 & N/A & N/A & 80.44 & N/A
\\
TransUNet~\cite{chen2021transunet} & 94.09 & 82.07 & 89.92 & 63.07 & 95.55 & 89.12 & 79.53 & N/A & N/A & 84.76 & N/A
\\
UNETR~\cite{hatamizadeh2022unetr} & 92.23 & 91.28 & 94.19 & 56.20 & 94.25 & 86.73 & 72.56 & 91.56 & 93.31 & 83.92 & 85.81
\\
Swin UNETR~\cite{tang2022self} & 93.51 & 66.34 & 90.63 & 61.05 & 94.73 & 87.37 & 73.77 & 93.72 & 92.17 & 81.05 & 83.69
\\
\hline
Universal Model & \textbf{95.76} & \textbf{94.99} & 94.42 & \textbf{88.79} & \textbf{97.03} & 89.36 & \textbf{80.99} & \textbf{97.71} & \textbf{96.72} & \textbf{91.62} & \textbf{92.86}
\\ \hline
\vspace{0.2 em}\\
\textbf{\textit{JHH}} & spleen & kidneyR & kidneyL
& gallbladder & liver & stomach 
& pancreas & arota & postcava
& vein & mDSC
\\ \shline
SegResNet~\cite{siddiquee2021redundancy} & 93.11 & 89.92 & 87.84 & 74.62 & 95.37 & 87.90 & 76.33 & 84.05 & 79.36 & 57.13 & 82.56
\\
nnFormer~\cite{zhou2021nnformer} & 86.71 & 87.03 & 84.28 & 63.37 & 91.64 & 73.18 & 71.88 & 84.73 & 78.61 & 55.31 & 77.67
\\
UNesT~\cite{yu2022unest} & 93.82 & 90.42 & 89.04 & 76.40 & 95.30 & 89.65 & 78.97 & 84.36 & 79.61 & 59.70 & 83.73
\\
TransBTS~\cite{wang2021transbts} & 85.47 & 81.58 & 82.00 & 60.58 & 92.50 & 72.29 & 63.25 & 83.47 & 75.07 & 55.38 & 75.16
\\
TransUNet~\cite{chen2021transunet} & 94.63 & 89.86 & 89.61 & 77.28 & 95.85 & 88.95 & 79.98 & 85.06 & \textbf{81.02} & \textbf{59.76} & 84.20
\\
UNETR~\cite{hatamizadeh2022unetr} & 91.89 & 89.07 & 87.60 & 66.97 & 91.48 & 83.18 & 70.56 & 82.92 & 75.20 & 57.53 & 79.64
\\
Swin UNETR~\cite{tang2022self} & 92.23 & 84.34 & 82.95 & 74.06 & 94.91 & 82.28 & 71.17 & \textbf{85.50} & 79.18 & 55.11 & 80.17
\\
\hline
Universal Model & \textbf{93.94} & \textbf{91.53} & \textbf{90.21} & \textbf{84.15} & \textbf{96.25} & \textbf{92.51} & \textbf{82.72} & 77.35 & 79.64 & 57.10 & \textbf{84.54}
\\ \hline
\end{tabular}%
\label{tab:generalizability}
\end{table*}

%%%%% tab:transfer_learning
\begin{table*}[h]
\caption%
{\textbf{\textit{Transferability:} Fine-tuning performance.} Fine-tuning Universal Model significantly outperforms learning from scratch on two downstream datasets (\ie, TotalSegmentator and JHH). Moreover, Universal Model, trained by image segmentation as proxy task, can extract better visual representation---more related to segmentation tasks---than other pre-trained models developed in the medical domain. Due to the space, the per-class evaluation of TotalSegmentator and JHH can be found in Appendix Tables~\ref{tab:TotalSeg_vertebrae}--\ref{tab:TotalSeg_organs} and \tableautorefname~\ref{tab:JHH_finetuning}, respectively.}
\centering
\footnotesize
\begin{tabular}{p{0.13\linewidth}|P{0.12\linewidth}P{0.12\linewidth}P{0.12\linewidth}P{0.12\linewidth}P{0.11\linewidth}P{0.11\linewidth}}
\hline
Method & TotalSeg\_vertebrae & TotalSeg\_cardiac & TotalSeg\_muscles & TotalSeg\_organs & JHH\_cardiac & JHH\_organs
\\ \shline
Scratch & 81.06&84.47&88.83&86.42 & 71.63 & 89.08
\\
MedicalNet~\cite{chen2019med3d}&82.28&87.40&91.36&86.90 & 58.07 & 77.68
\\
Models Gen.~\cite{zhou2019models}&85.12&86.51&89.96&85.78
& \textbf{74.25} & 88.64
\\
Swin UNETR~\cite{tang2022self}&86.23&87.91&92.39&88.56
& 67.85 & 87.21
\\
UniMiSS~\cite{xie2022unimiss}&85.12&88.96&92.86&88.51 & 69.33 & 82.53
\\
\hline
Universal Model & \textbf{86.49} & \textbf{89.57} & \textbf{94.43} & \textbf{88.95} & 72.06 & \textbf{89.37}
\\
\hline
\end{tabular}%
\label{tab:transfer_learning}
% \vspace{-0.2cm}
\end{table*}

\section{Intriguing Properties}
\label{sec:properties}

\noindent\textbf{\textit{Efficiency: FLOPs vs. DSC.}}
It is clinically important to make AI models faster~\cite{chen2019augmented,esteva2021deep}. The floating-point operations per second (FLOPS) are used to indicate the inference speed. 
\figureautorefname~\ref{fig:computational_efficiency} presents a speed-performance plot, showing that Universal Model is computationally more efficient compared with dataset-specific models ($>$6$\times$ faster), while maintaining a high DSC score of 74\% on average\footnote{Existing dataset-specific models are limited to being trained individually for each MSD task, due to the partial label problem.}.  

\smallskip\noindent\textbf{\textit{Expansibility: flexible backbones.}}
The proposed Universal Model framework can be applied flexibly to other backbones. We further conduct experiments in CNN-based backbone (\ie, U-Net~\cite{ronneberger2015u}) and achieve an average DSC score of 76.73\% over 25 organs and 6 tumors, which is comparable with the average DSC score of 76.11\% obtained by Swin UNETR, as shown in \tableautorefname~\ref{fig:expansibility}. 

\smallskip\noindent\textbf{\textit{Generalizability: results on external datasets.}}
A key expectation of medical AI models is their generalizability, \ie, performance on new data across many hospitals, rather than the performance tailored to a single dataset~\cite{mongan2020checklist,norgeot2020minimum,hu2023label}. Compared with dataset-specific models, Universal Model has trained on the order of magnitude more diverse CT scans, therefore demonstrating significantly better generalizability. We conduct the evaluation on 3D-IRCADb (public) and JHH (private), which are absolutely not seen in the training. Universal Model substantially outperforms the previous methods on 3D-IRCADb and JHH with a DSC improvement of 5\% and 4\%, respectively (see \tableautorefname~\ref{tab:generalizability}).

\smallskip\noindent\textbf{\textit{Transferability: fine-tuning results.}}
Universal Model can serve as a powerful pre-training model for segmentation. Through pre-training by assembly dataset directly and fine-tuning to other datasets, the Universal Model achieves the highest DSC compared with other pre-training methods with 86.49\%, 89.57\%, 94.43\% and 88.95\% for four downstream tasks in the TotalSegmentator dataset (see \tableautorefname~\ref{tab:transfer_learning}). 
% This demonstrates the potential for improving the generalization of medical imaging model embedding by directly capturing the fine-grained information for segmentation.
%{\jlred Since the other unrelated pre-training tasks (reconstruction, colorization, jigsaw) can not capture the fine-grained information for segmentation.}

\section{Conclusion}
\label{sec:conclusion}

This work presents a CLIP-Driven Universal Model for abdominal organ segmentation and tumor detection. To address the label inconsistency and orthogonality problems, we integrate CLIP embedding with segmentation models, resulting in a flexible and powerful segmentor. The model can effectively learn from partially labeled datasets and achieve high performance, as evidenced by ranking first in both MSD and BTCV. The segmentation accuracy of six organs has approached that of humans. Importantly, our study demonstrates that CLIP embedding can establish a stronger and more meaningful anatomical relationship between organs and tumors than the widely-used one-hot embedding as the ground truth. Furthermore, we validate several clinically important merits of the CLIP-Driven Universal Model, including compelling efficiency, generalizability, transferability, and expansibility, through experimental results.

\smallskip\noindent\textbf{Acknowledgments.}
This work was supported by the Lustgarten Foundation for Pancreatic Cancer Research, the Patrick J. McGovern Foundation Award, and the National Natural Science Foundation of China (62001410). We thank Ali Hatamizadeh, Tong Li, and Wenxuan Li for their constructive suggestions at several stages of the project.

\clearpage
{\small
\bibliographystyle{ieee_fullname}
\bibliography{refs,zzhou}
}

\newcolumntype{x}[1]{>{\centering\arraybackslash}p{#1pt}}
\newcolumntype{z}[1]{>{\raggedright\arraybackslash}p{#1pt}}

\newpage
\clearpage

\begin{table*}[t]
\caption{
\textbf{The information for an assembly of datasets.} We have developed a \textit{Universal Model} from an assembly of 1--14 public datasets. The official test and validation sets of Medical Segmentation Decathlon (MSD) and Beyond the Cranial Vault (BTCV) are used to benchmark the performance of organ segmentation (\S\ref{sec:strong_challenge_ranking}) and tumor detection (\S\ref{sec:high_specificity}). 3D-IRCADb (15), TotalSegmentator (16) and a large-scale private dataset (17), consisting of 5,038 CT scans with 21 annotated organs, are used for independent evaluation of model generalizability and transferability (\S\ref{sec:properties}). This list will continue to grow when more annotated datasets become available.
}
\centering
\footnotesize
\begin{tabular}{p{0.158\linewidth}P{0.06\linewidth}P{0.05\linewidth}p{0.637\linewidth}}
    \hline
     Datasets & \# Targets & \# Scans & Annotated Organs or Tumors \\
    \shline
    1. Pancreas-CT \cite{roth2015deeporgan} & 1 & 82 & {Pancreas} \\
    2. LiTS \cite{bilic2019liver} & 2 & 201 & {Liver, Liver Tumor$^*$} \\
    3. KiTS \cite{heller2020international} & 2 & 300 & {Kidney, Kidney Tumor$^*$} \\
    4. AbdomenCT-1K \cite{ma2021abdomenct} & 4 & 1,000 & {Spleen, Kidney, Liver, Pancreas} \\
    5. CT-ORG \cite{rister2020ct} & 4 & 140 & {Lung, Liver, Kidneys and Bladder} \\
    6. CHAOS \cite{valindria2018multi} & 4 & 40 & {Liver, Left Kidney, Right Kidney, Spl} \\
    7-11. MSD CT Tasks \cite{antonelli2021medical} & 9 & 947 & {Spl, Liver and Tumor$^*$, Lung Tumor$^*$, Colon Tumor$^*$, Pan and Tumor$^*$, Hepatic Vessel and Tumor$^*$} \\
    12. BTCV \cite{landman2015miccai} & 13 & 50 & {Spl, RKid, LKid, Gall, Eso, Liv, Sto, Aor, IVC, R\&SVeins, Pan, RAG, LAG} \\
    13. AMOS22 \cite{ji2022amos} & 15 & 500 & {Spl, RKid, LKid, Gall, Eso, Liv, Sto, Aor, IVC, Pan, RAG, LAG, Duo, Bla, Pro/UTE} \\
    14. WORD \cite{luo2021word} & 16 & 150 & {Spl, RKid, LKid, Gall, Eso, Liv, Sto, Pan, RAG, Duo, Col, Int, Rec, Bla, LFH, RFH} \\
    \midrule
    15. 3D-IRCADb \cite{soler20103d} & 13 & 20 & {Liv, Liv Cyst, RLung, LLung, Venous, PVein, Aor, Spl, RKid, LKid, Gall, IVC} \\
    \cmidrule{4-4}
    \multirow{5}{*}{16. TotalSegmentator \cite{wasserthal2022totalsegmentator}}  & \multirow{5}{*}{104} & \multirow{5}{*}{1,024} & { Clavicula, Humerus, Scapula, Rib 1-12, Vertebrae C1-7, Vertebrae T1-9, Vertebrae L1-5, Hip, Sacrum, Femur, Aorta, Pulmonary Artery, Right Ventricle, Right Atrium, Left Atrium, Left Ventricle, Myocardium, PVein, SVein, IVC, Iliac Artery, Iliac Vena, Brain, Trachea, Lung Upper Lobe, Lung Middle Lobe, Lung Lower Lobe, AG, Spl, Liv, Gall, Pan, Kid, Eso, Sto, Duo, Small Bowel, Colon, Bla, Autochthon, Iliopsoas, Gluteus Minimus, Gluteus Medius, Gluteus Maximus} \\
    \cmidrule{4-4}
    \multirow{3}{*}{17. JHH (\textit{private})} & \multirow{3}{*}{21} & \multirow{3}{*}{5,038} & Aor, AG, CBD, Celiac AA, Colon, duo, Gall, IVC, Lkid, RKid, Liv, Pan, Pan Duct, SMA, Small bowel, Spl, Sto, Veins, Kid LtRV, Kid RtRV, CBD Stent, PDAC$^*$, PanNET$^*$, Pancreatic Cyst$^*$\\
    \hline
\end{tabular}
\label{tab:public_dataset}
\end{table*}

\section*{Appendix: CLIP-Driven Universal Model}
\appendix

\noindent\textbf{Abstract.} In this supplementary material, we provide additional information about the CLIP-Driven Universal Model and the assembly of 14 public datasets, as well as more detailed experimental results than those in the main paper. 
%Appendix~\ref{sec:novelty_analysis} offers the novelty analysis of the proposed CLIP-Driven Universal Model. 
Appendix~\ref{sec:medical_prompt_template} discusses the influence of the medical prompt template. Appendix~\ref{sec:assembly_datasets} provides the specifications for the assembly of datasets. Appendix~\ref{sec:implementation_details} elaborates on the implementation details, including the data augmentations, model network structures and evaluation metrics used in the main paper. 
%In Appendix~\ref{sec:evaluation_metrics}, we define the evaluation metrics used in the main paper. 
Appendix~\ref{sec:additional_evaluation} supplements the qualitative and quantitative analysis in the main paper, including the visualization of kidney tumors and liver tumors, complete evaluation results of the transfer learning experiment, and whole embedding space visualization. Finally, Appendix~\ref{sec:open_challenges_review} visualizes several open challenges when assembling public datasets with partial labels.
% discussed in \S\ref{sec::assembly_of_public_dataset} of the main paper. 

\section{Medical Prompt Template}
\label{sec:medical_prompt_template}

To fully explore the effect of templates on CLIP embedding, an experiment is performed in the whole assembly of datasets as shown in \tableautorefname~\ref{tab:clip_embedding}.
Four text templates are employed to show the context, \ie, ``V1: A computerized tomography of a [CLS].'', ``V2: There is [CLS] in this computerized tomography.'', ``V3: This computerized tomography has a [CLS].'', ``V4: A photo of a  [CLS].''. 
The effectiveness of the prompt template is slightly different from the toy experiment. With increasing organ numbers, templates V1 and V2 still show better performance in encoding the relationship, but template V3 would deteriorate the results. In addition, a widely used template V4 could also promote the segmentation performance. 

As known, the prompt template is a crucial factor for text model \cite{zhou2022learning,ma2021template}. How select an appropriate template is still an open problem for the medical image text-vision models. We encourage more future work to explore this area.

\section{Assembly of Datasets}
\label{sec:assembly_datasets}

The assembly of datasets consists of 14 publicly available datasets for training and 2 public datasets and 1 large-scale private dataset for testing (summarized in~\tableautorefname~\ref{tab:public_dataset}).
It is non-trivial to assemble datasets annotated from various institutions since the annotation protocols are inconsistent. As mentioned in the main paper, we unify the label index for all datasets. The corresponding relationship is as follows. (Spleen, 1); (Right Kidney, 2); (Left Kidney, 3); (Gall Bladder, 4); (Esophagus, 5); (Liver, 6); (Stomach, 7); (Aorta, 8); (Postcava, 9); (Portal Vein and Splenic Vein, 10); (Pancreas, 11); (Right Adrenal Gland, 12); (Left Adrenal Gland, 13); (Duodenum, 14); (Hepatic Vessel, 15); (Right Lung, 16); (Left Lung, 17); (Colon, 18); (Intestine, 19); (Rectum, 20); (Bladder, 21); (Prostate/Uterus, 22); (Head of Femur Left, 23); (Head of Femur Right, 24); (Celiac Truck, 25); (Kidney Tumor, 26); (Liver Tumor, 27); (Pancreas Tumor, 28); (Hepatic Vessel Tumor, 29); (Lung Tumor, 30); (Colon Tumor, 31); (Kidney Cyst, 32). 
Firstly, we map all the datasets into the standard index template. Then, for these datasets (KiTS, WORD, AbdomenCT-1K, and CT-ORG), which do not distinguish between the left and right organs, we split the organ (Kidney, Adrenal Gland, and Lung) into left part and right part through the script. In addition, we have taken the inclusion relation into consideration, \eg, the organ tumor is part of the organ, and the hepatic vessel is inside the liver. Since we formulate each organ segmentation result as a binary mask, we can organize the segmentation ground truth for these overlapped organs independently in a binary mask manner.

\smallskip\noindent\textbf{Pancreas-CT}~\cite{roth2015deeporgan} consists of 82 contrast-enhanced abdominal CT volumes. This dataset only provides the pancreas label annotated by an experienced radiologist, and all CT scans have no pancreatic tumor.

\smallskip\noindent\textbf{LiTS}~\cite{bilic2019liver} contains 131 and 70 contrast-enhanced 3-D abdominal CT scans for training and testing, respectively. The data set was acquired by different scanners and protocols at six different clinical sites, with a largely varying in-plane resolution from 0.55 to 1.0 mm and slice spacing from 0.45 to 6.0 mm.

\smallskip\noindent\textbf{KiTS}~\cite{heller2020international} includes 210 training cases and 90 testing cases with annotations provided by the University of Minnesota Medical Center. Each CT scan has one or more kidney
tumors.

\smallskip\noindent\textbf{AbdomenCT-1K}~\cite{ma2021abdomenct} consists of 1112 CT scans from five datasets with liver, kidney, spleen, and pancreas annotations.

\smallskip\noindent\textbf{CT-ORG}~\cite{rister2020ct} is composed of 140 CT images containing 6 organ classes, which are from 8 different medical centers. Most of the images exhibit liver lesions, both benign and malignant.

\smallskip\noindent\textbf{CHAOS}~\cite{valindria2018multi} provides 20 patients for multi-organ segmentation. All CT scans have no liver tumor.

\smallskip\noindent\textbf{MSD CT Tasks}~\cite{antonelli2021medical} includes liver, lung, pancreas, colon, hepatic vessel, and spleen tasks for a total of 947 CT scans with 4 organs and 5 tumors.

\smallskip\noindent\textbf{BTCV}~\cite{landman2015miccai} consists of 50 abdominal CT scans from metastatic liver cancer patients or post-operative ventral hernia patients. They are collected from the Vanderbilt University Medical Center.

\smallskip\noindent\textbf{AMOS22}~\cite{ji2022amos} is the abbreviation of the multi-modality abdominal multi-organ segmentation challenge of 2022. The AMOS dataset contains 500 CT with voxel-level annotations of 15 abdominal organs.

\smallskip\noindent\textbf{WORD}~\cite{luo2021word} collects 150 CT scans from 150 patients before the radiation therapy in a single center. All of them are scanned by a SIEMENS CT scanner without appearance enhancement. Each CT volume consists of 159 to 330 slices of 512 × 512 pixels.

\smallskip\noindent\textbf{3D-IRCADb}~\cite{soler20103d} contains 20 venous phase enhanced CT scans. Each CT scan has various annotations, and only annotated organs are tested to validate the model's generalizability.

\smallskip\noindent\textbf{TotalSegmentator}~\cite{wasserthal2022totalsegmentator} collects 1024 CT scans randomly sampled from PACS over the timespan of the last 10 years. The dataset contains CT images with different sequences (native, arterial, portal venous, late phase, dual-energy), with and without contrast agent, with different bulb voltages, with different slice thicknesses and resolution and with different kernels (soft tissue kernel, bone kernel).

\smallskip\noindent\textbf{JHH (\textit{private})} contains 5038 CT scans with 21 annotated organs, where each case was scanned by contrast-enhanced CT in both venous and arterial phases, acquired on Siemens MDCT scanners. The JHH dataset is used to investigate the extensibility of new classes.

% \newpage
\section{Implementation Details}
\label{sec:implementation_details}

\subsection{Data Augmentation}
Our data augmentation is implemented in python with MONAI\footnote{\href{https://monai.io/}{https://monai.io/}}. The orientation of CT scans is changed into specified axcodes. Isotropic spacing is adopted to re-slice each scan to the same voxel size of $1.5 \times 1.5 \times 1.5mm^3$. We truncate the intensity in each scan to the range $[-175, 250]$ and linearly normalize them to $[0, 1]$. Considering the valid part is part of the whole medical image, we crop only the foreground object based on the images. During training, we crop random fixed-sized $96 \times 96 \times 96$ regions with the center being a foreground or background voxel based on the predefined ratio. Also, we randomly rotate the input patch by 90 degrees and shift intensity with $0.1$ offset with $0.1$ and $0.2$ probability. To avoid confusion between the organ in the right and left parts, we do not use mirroring augmentation.

\begin{table}[t]
\caption{
    \textbf{The 5-fold cross-validation performance on MSD.} These are the tabular comparison between Universal Model and Swin UNETR~\cite{tang2022self} (previously ranked first on the MSD leaderboard).
    The performance is evaluated by DSC scores.
    % We compare  on 5-hold cross validation of the MSD dataset.
}
\centering
\footnotesize
\begin{tabular}{p{0.12\linewidth}p{0.18\linewidth}|P{0.25\linewidth}P{0.24\linewidth}}
\hline
\multicolumn{2}{l|}{Task} & SwinUNETR \cite{tang2022self} & Ours
\\
\shline
\multirow{2}{*}{Task 03} & Liver & 94.12$\pm$2.34 & \textbf{96.49}$\pm$\textbf{0.23}
\\
% \cline{2-4}
& Liver Tumor & 57.86$\pm$4.72 & \textbf{71.94$\pm$3.74}
\\
\hline
Task 06 & Lung Tmuor  & \textbf{68.90$\pm$5.44} & 67.15$\pm$5.81
\\ 
\hline
\multirow{2}{*}{Task 07} & Pancreas & 80.06$\pm$0.83 & \textbf{82.70$\pm$1.96}
\\
% \cline{2-4}
& Panc. Tumor & 52.53$\pm$3.76 & \textbf{60.82$\pm$10.2} 
\\
\hline
\multirow{2}{*}{Task 08} & Hepat. Ves. & 62.33$\pm$2.44 & \textbf{62.55$\pm$3.64}
\\
% \cline{2-4}
& Ves. Tumor & 68.56$\pm$3.82 & \textbf{69.39$\pm$2.29}
\\
\hline
Task 09 & Spleen & 95.80$\pm$0.56 & \textbf{96.71$\pm$0.21}
\\ 
\hline
Task 10 & Col. Tumor & 50.45$\pm$10.1 & \textbf{62.14$\pm$17.8}
\\ 
\hline
\end{tabular}
\label{tab:msd_testset}
\end{table}

\subsection{Network Structures}
\noindent\textbf{Text branch.}
We apply the pre-trained text encoder ``ViT-B/32'' of the CLIP as the text branch\footnote{\href{https://github.com/openai/CLIP}{https://github.com/openai/CLIP}}.
We can extract and store the text features to reduce overhead brought by the text encoder in the training and inference stage since the CLIP embedding only depends on the dictionary, which is fixed.

\smallskip\noindent\textbf{Vision branch.}
We adopt Swin UNETR as a vision encoder. The Swin UNETR consists of 4 attention stages comprising 2 transformer blocks and 5 convolution stages comprising of CNN-based structure. In the attention stage, a patch merging layer is used to reduce the resolution by a factor of 2. Stage 1 consists of a linear embedding layer and transformer blocks that maintain the number of tokens as $\frac{H}{2} \times \frac{W}{2} \times \frac{D}{2}$. a patch merging layer groups patches with resolution $2 \times 2 \times 2$ and concatenates them, resulting in a 4C-dimensional feature embedding. A linear layer is then used to down-sample the resolution by reducing the dimension to 2C. The same procedure continues in stages 2, 3, and 4 \cite{tang2022self}. 
The text-based controller is a single convolutional layer, which takes the CLIP embedding and global pooling feature from the last convolution stages in the vision encoder as input.

\subsection{Evaluation Metrics}
\label{sec:evaluation_metrics}
The Dice similarity coefficient (DSC) and Normalized Surface Distance (NSD) are used as measurements for 3D segmentation results. The DSC metric is defined as:
\begin{equation}
    \text{DSC}=\frac{2 \sum_{i=1}^I Y_i \hat{Y}_i}{\sum_{i=1}^I Y_i+\sum_{i=1}^I \hat{Y}_i},
\end{equation}
where $Y$ and $\hat{Y}$ denote the ground truth and prediction of voxel values. The details of Normalized Surface Distance (NSD) could refer to Sec. 4.6 in~\cite{nikolov2018deep}.

%%%%% fig:pseudo_truth_visualization
\begin{figure*}[t]
\centerline{\includegraphics[width=1.0\linewidth]{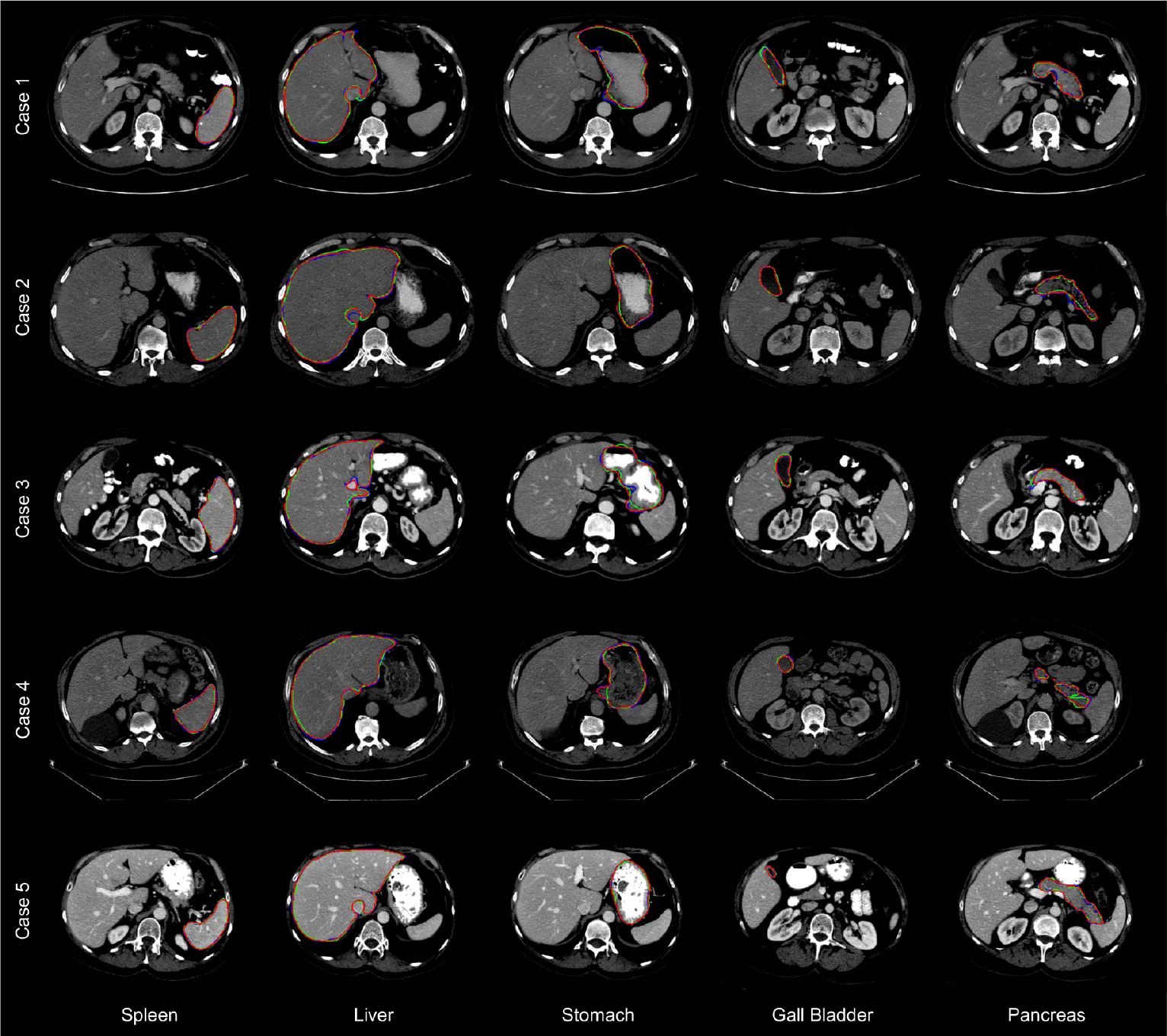}}
\caption{
\textbf{Contour line comparison among pseudo labels and two human experts.} The \textcolor{red}{red} line represents the annotation from Doctor 1; \textcolor{green}{green} line indicates the annotation from Doctor 2; \textcolor{blue}{blue} line shows the results generated by Universal Model. Examples of CT scans annotated by our pseudo labels and two human experts with contour line comparison. The prediction results of these organs generated by the medical model are comparable with human experts.
}
\label{fig:pseudo_truth_visualization}
\end{figure*}

%%%%%% fig:qualitative_visualization_liver_tumor
\begin{figure*}[t]
\centerline{\includegraphics[width=1.0\linewidth]{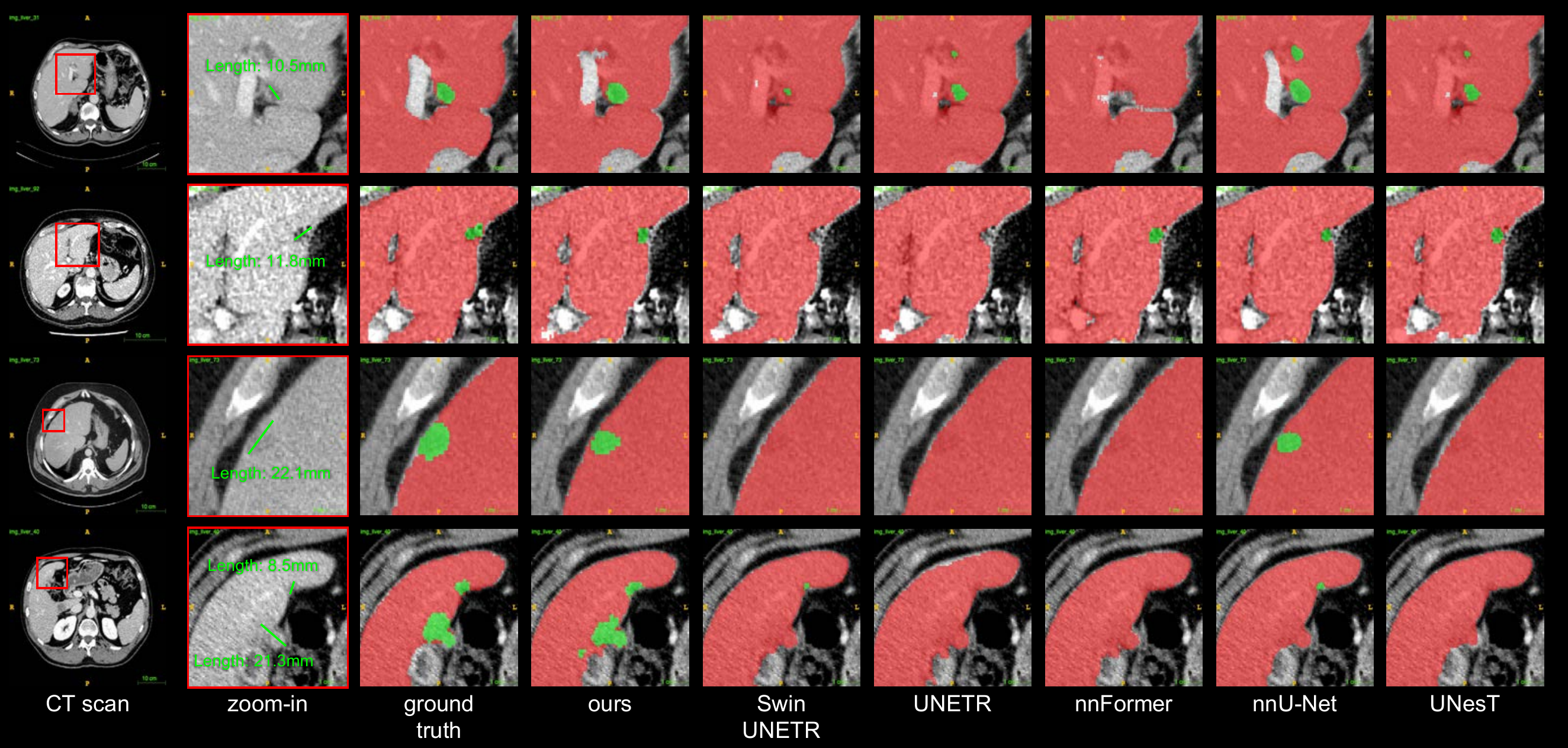}}
\caption{
\textbf{Liver tumor detection.} Qualitative visualizations of the proposed Universal Model and four competitive baseline methods. We review the detection results of tumors from smaller to larger sizes (Rows 1--4). The Universal Model succeeds in detecting small tumors ignored by other methods and in detecting multiple tumors in one CT. In addition, it avoids the false positive prediction, which validates the good practicability of Universal Model. 
}
\label{fig:qualitative_visualization_liver_tumor}
\end{figure*}

%%%%% fig:qualitative_visualization_kidney_tumor
\begin{figure*}[t]
\centerline{\includegraphics[width=1.0\linewidth]{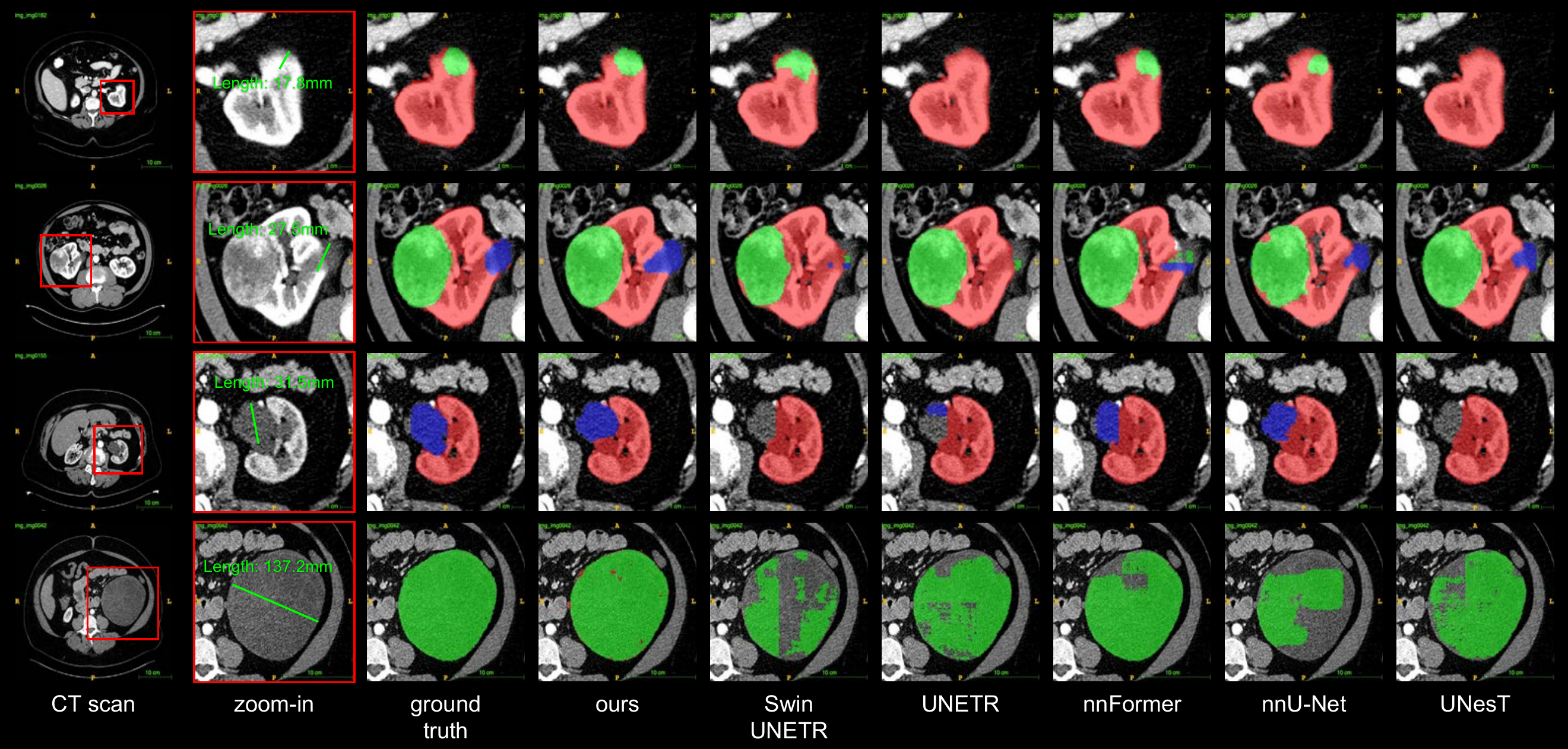}}
\caption{
\textbf{Kidney tumor detection.} Qualitative visualizations of the proposed Universal Model and four competitive baseline methods. We review the detection results of tumors from smaller to larger sizes (Rows 1--4). The Universal Model can detect well not only on the kidneys ({\jlred red} region), but also kidney tumors ({\jlgreen green} region) and cysts ({\jlblue blue} region).
}
\label{fig:qualitative_visualization_kidney_tumor}
\end{figure*}

%%%%%%% TotalSeg_vertebrae
\begin{table*}[t]
\caption
{\textbf{The complete evaluation of TotalSeg\_vertebrae.} The results are evaluated by DSC. Our Universal Model represents the best transferability.}
\centering
\footnotesize
\begin{tabular}{p{0.13\linewidth}|P{0.04\linewidth}P{0.04\linewidth}P{0.04\linewidth}P{0.04\linewidth}P{0.04\linewidth}P{0.04\linewidth}P{0.04\linewidth}P{0.04\linewidth}P{0.04\linewidth}P{0.04\linewidth}P{0.04\linewidth}P{0.04\linewidth}P{0.05\linewidth}}
\hline
\textbf{\textit{Method}} & L5&L4&L3&L2&L1&T12&T11&T10&T9&T8&T7&T6
\\ \shline
Scratch & 86.68 & 88.37 & 89.83 & 84.28 & 91.98 & 87.45 & 88.29 & 86.78 & 83.50 & 75.70 & 77.73 & 75.84
\\
MedicalNet~\cite{chen2019med3d}& \textbf{91.72} & 91.01 & 86.03 & 84.73 & 91.52 & \textbf{89.98} & 89.06 & 89.35 & 85.71 & 82.99 & 81.54 & 79.74
\\
Models Gen.~\cite{zhou2019models}&89.64 & 89.24 & 89.38 & 82.85 & 90.79 & 88.62 & 90.11 & 90.43 & 89.22 & \textbf{85.21} & 80.83 & 77.40
\\
Swin UNETR~\cite{tang2022self}&89.56 & 90.80 & 93.08 & 86.38 & \textbf{94.35} & 89.65 & \textbf{92.02} & \textbf{91.99} & \textbf{89.65} & 82.20 & 85.01 & \textbf{81.06}
\\
UniMiSS~\cite{xie2022unimiss}& 89.20 & 91.21 & \textbf{94.16} & 86.61 & 91.57 & 87.29 & 90.18 & 90.56 & 88.09 & 83.47 & 80.73 & 76.40
\\
\hline
Universal Model&88.95 & \textbf{91.38} & 93.82 & \textbf{87.04} & 93.53 & 88.96 & 90.50 & 91.40 & 89.18 & 84.25 & \textbf{83.63} & 79.95
\\ \hline
\vspace{1.0 em}\\
\textbf{\textit{Method}}&T5&T4&T3&T2&T1&C7&C6&C5&C4&C3&C2&C1&Average
\\ \shline
Scratch & 73.14 & 72.26 & 77.12 & 80.36 & 85.76 & 83.39 & 69.80 & 70.23 & 69.82 & 85.74 & 83.35 & 78.18 & 81.06
\\
MedicalNet~\cite{chen2019med3d}&77.28 & 76.60 & 76.57 & 80.94 & 85.54 & 83.05 & 76.05 & 73.04 & 80.55 & 74.35 & 74.67 & 72.91 & 82.28
\\
Models Gen.~\cite{zhou2019models}&79.59 & \textbf{78.73} & 82.01 & 84.63 & 90.02 & 88.20 & \textbf{81.09} & 78.90 & 78.21 & 89.69 & 88.06 & 80.23 & 85.12
\\
Swin UNETR~\cite{tang2022self}&82.33 & 77.74 & 81.78 & 83.53 & 88.22 & 87.81 & 78.38 & 80.36 & 83.00 & 92.68 & 87.97 & 80.16 & 86.23
\\
UniMiSS~\cite{xie2022unimiss}&78.97 & 76.60 & 82.33 & 85.14 & 90.04 & 88.68 & 79.18 & 79.17 & 79.00 & 88.19 & 86.38 & 79.80 & 85.12
\\
\hline
Universal Model& \textbf{83.07} & 78.67 & \textbf{82.97} & \textbf{86.06} & \textbf{90.67} & \textbf{88.75} & 77.03 & \textbf{80.87} & \textbf{83.05} & \textbf{92.94} & \textbf{88.20} & \textbf{80.87} & \textbf{86.49}
\\ \hline
\end{tabular}%
\label{tab:TotalSeg_vertebrae}
\end{table*}

%%%%%%% TotalSeg_cardiac
\begin{table*}[t]
\caption[caption]%
{\textbf{The complete evaluation of TotalSeg\_cardiac.} The results are evaluated by DSC. Our Universal Model represents the best transferability. The abbreviation in the table is listed as follows. HM (heart myocardium), HA (heart atrium), HV (heart ventricle), PA (pulmonary artery), IA (iliac artery), IV (iliac vena), UB (urinary bladder).}
\centering
\footnotesize
\begin{tabular}{p{0.13\linewidth}|P{0.06\linewidth}P{0.06\linewidth}P{0.06\linewidth}P{0.06\linewidth}P{0.06\linewidth}P{0.06\linewidth}P{0.06\linewidth}P{0.06\linewidth}P{0.06\linewidth}P{0.06\linewidth}}
\hline
\textbf{\textit{Method}}&esophagus & trachea & HM & HA\_left & HV\_left & HA\_right & HV\_right & PA & brain
\\ \shline
Scratch & 84.73 & 90.72 & 85.53 & 91.78 & 91.15 & 90.10 & 88.25 & 87.20 & 93.79
\\
MedicalNet~\cite{chen2019med3d}&89.43 & 94.08 & 88.71 & 93.50 & 92.17 & 90.90 & 90.83 & 89.51 & 95.11
\\
Models Gen.~\cite{zhou2019models}&87.96 & 93.47 & 87.40 & 93.61 & 92.23 & 92.02 & 89.74 & 89.34 & 94.99
\\
Swin UNETR~\cite{tang2022self}&89.77 & 94.37 & 88.85 & 94.42 & 92.99 & 92.61 & 90.40 & 88.91 & 95.14
\\
UniMiSS~\cite{xie2022unimiss}& 90.45 & 94.51 & 90.29 & 94.34 & 93.70 & 93.10 & 91.46 & 89.67 & 94.99
\\
\hline
Universal Model& \textbf{90.97} & \textbf{94.71} & \textbf{90.88} & \textbf{94.64} & \textbf{93.72} & \textbf{93.30} & \textbf{91.66} & \textbf{90.80} & \textbf{95.34}
\\ \hline
\vspace{1.0 em}\\
\textbf{\textit{Method}}& IA\_left & IA\_right & IV\_left & IV\_right & small\_bow. & duodenum & colon & UB & face & Average
\\ \shline
Scratch & 80.32 & 79.78 & 79.80 & 81.69 & 81.97 & 72.21 & 82.51 & 89.59 & 69.40 & 84.47
\\
MedicalNet~\cite{chen2019med3d}&87.06 & 84.90 & 86.93 & 86.46 & 83.14 & 72.01 & 84.22 & 90.43 & 73.85 & 87.40
\\
Models Gen.~\cite{zhou2019models}&85.71 & 83.09 & 85.77 & 85.79 & 81.75 & 69.37 & 85.25 & 90.31 & 69.42 & 86.51
\\
Swin UNETR~\cite{tang2022self}&88.26 & 86.44 & 87.13 & 87.59 & 83.29 & 70.71 & 87.50 & 89.93 & 74.08 & 87.91
\\
UniMiSS~\cite{xie2022unimiss}&89.18 & 87.81 & 89.04 & 88.55 & 84.83 & 74.74 & 88.16 & 91.83 & 74.76 & 88.96
\\
\hline
Universal Model& \textbf{89.89} & \textbf{88.54} & \textbf{89.58} & \textbf{89.27} & \textbf{84.85} & \textbf{76.23} & \textbf{89.06} & \textbf{92.07} & \textbf{76.81} & \textbf{89.57}
\\ \hline
\end{tabular}%
\label{tab:TotalSeg_cardiac}
\end{table*}

%%%%%%% TotalSeg_muscles
\begin{table*}[t]
\caption
{\textbf{The complete evaluation of TotalSeg\_muscles.} The results are evaluated by DSC. Our Universal Model represents the best transferability. The abbreviation in the table is listed as follows. Clav. (Clavicula), GMa (gluteus maximus), GMe (gluteus medius), GMi (gluteus minimus), Aotu. (Autochthon)}
\centering
\footnotesize
\begin{tabular}{p{0.12\linewidth}|P{0.06\linewidth}P{0.06\linewidth}P{0.06\linewidth}P{0.06\linewidth}P{0.044\linewidth}P{0.044\linewidth}P{0.044\linewidth}P{0.044\linewidth}P{0.06\linewidth}P{0.06\linewidth}P{0.052\linewidth}}
\hline
\textbf{\textit{Method}}& Humerus\_L & Humerus\_R & Scapula\_L & Scapula\_R & Clav.\_L & Clav.\_R & Femur\_L & Femur\_R & Hip\_L & Hip\_R & Sacrum
\\ \shline
Scratch & 84.27 & 84.44 & 91.71 & 89.78 & 80.38 & 75.81 & 93.41 & 93.02 & 92.90 & 88.66 & 83.63
\\
MedicalNet~\cite{chen2019med3d}& 87.25 & 85.67 & 88.68 & 92.62 & 94.35 & 93.96 & 84.85 & 96.59 & 96.98 & 96.31 & 95.19
\\
Models Gen.~\cite{zhou2019models}& 90.61 & 79.73 & 88.56 & 92.06 & 91.19 & 92.57 & 86.08 & 93.57 & 85.35 & 82.40 & 87.91
\\
Swin UNETR~\cite{tang2022self}& 88.32 & 86.35 & 90.82 & 93.88 & 94.90 & 94.52 & 85.92 & 97.71 & 97.42 & 97.49 & 95.73
\\
UniMiSS~\cite{xie2022unimiss}& 89.73 & 92.30 & 91.72 & 94.77 & 94.57 & 93.66 & 84.92 & 97.67 & 97.35 & 97.11 & 96.18
\\
\hline
Universal Model & \textbf{91.32} & \textbf{93.87} & \textbf{93.11} & \textbf{95.59} & \textbf{95.00} & \textbf{95.88} & \textbf{86.79} & \textbf{98.48} & \textbf{98.04} & \textbf{98.32} & \textbf{96.94}
\\ \hline
\vspace{1.0 em}\\
\textbf{\textit{Method}}& GMa\_L & GMa\_R & GMe\_L & GMe\_R & GMi\_L & GMi\_R & Aotu.\_L & Aotu.\_R & Iliopsoas\_L & Iliopsoas\_R & Average
\\ \shline
Scratch & 95.53 & 91.78 & 85.27 & 94.80 & 86.54 & 93.01 & 95.17 & 93.44 & \textbf{87.99} & 83.95 & 88.83
\\
MedicalNet~\cite{chen2019med3d}& 94.69 & 95.72 & 92.17 & 89.15 & 89.76 & 90.77 & 94.45 & 94.24 & 80.29 & 84.94 & 91.36
\\
Models Gen.~\cite{zhou2019models}& 96.19 & 92.06 & 90.07 & \textbf{94.99} & 92.12 & 92.60 & 95.86 & 95.93 & 85.64 & 83.82 & 89.96
\\
Swin UNETR~\cite{tang2022self}& 95.32 & 96.34 & 93.57 & 89.87 & 90.75 & 91.74 & 95.16 & 94.86 & 83.53 & 86.00 & 92.39
\\
UniMiSS~\cite{xie2022unimiss}& 95.53 & 96.37 & 93.80 & 90.28 & 90.87 & 93.02 & 95.17 & 95.48 & 85.71 & 84.02 & 92.86
\\
\hline
Universal Model & \textbf{96.68} & \textbf{96.99} & \textbf{95.55} & 91.36 & \textbf{93.19} & \textbf{94.52} & \textbf{96.31} & \textbf{96.34} & 86.92 & \textbf{88.89} & \textbf{94.29}
\\ \hline
\end{tabular}%
\label{tab:TotalSeg_muscles}
\end{table*}

%%%%%%% TotalSeg_organs
\begin{table*}[t]
\caption
{\textbf{The complete evaluation of TotalSeg\_organs.} The results are evaluated by DSC. Our Universal Model represents the best transferability. The abbreviation in the table is listed as follows. IVC (inferior vena cava), PSV (portal vein and splenic vein), AG (adrenal gland), LUL (lung upper lobe), LLL (lung lower lobe), LML (lung middle lobe)}
\centering
\footnotesize
\begin{tabular}{p{0.13\linewidth}|P{0.07\linewidth}P{0.07\linewidth}P{0.07\linewidth}P{0.07\linewidth}P{0.07\linewidth}P{0.07\linewidth}P{0.07\linewidth}P{0.07\linewidth}P{0.07\linewidth}}
\hline
\textbf{\textit{Method}}& spleen & Kidney\_R & Kidney\_L & gallbladder & liver & stomach & aorta & IVC & PSV
\\ \shline
Scratch & 93.58 & 94.09 & 87.73 & 73.86 & 96.79 & 89.17 & 90.68 & 82.10 & 71.35
\\
MedicalNet~\cite{chen2019med3d}& 95.54 & 92.43 & 90.86 & 79.36 & 97.10 & 91.53 & 90.12 & 86.18 & 73.34
\\
Models Gen.~\cite{zhou2019models}& 95.60 & 94.37 & 88.51 & 78.39 & 97.39 & 91.68 & 93.18 & 85.94 & 74.58
\\
Swin UNETR~\cite{tang2022self}&89.77 & 94.37 & 88.85 & 74.42 & 92.99 & 92.61 & 90.40 & \textbf{88.91} & 75.14
\\
UniMiSS~\cite{xie2022unimiss}& 95.78 & \textbf{94.75} & 89.35 & 79.14 & 97.39 & 91.87 & \textbf{93.50} & 86.19 & 75.26
\\
\hline
Universal Model & \textbf{96.24} & 94.67 & \textbf{91.43} & \textbf{81.48} & \textbf{97.63} & \textbf{92.76} & 92.22 & 87.87 & \textbf{76.10}
\\ \hline
\vspace{1.0 em}\\
\textbf{\textit{Method}} & pancreas & AG\_R & AG\_L & LUL\_L & LLL\_L & LUL\_R & LML\_R & LLL\_R & Average
\\ \shline
Scratch & 80.80 & 78.94 & 72.83 & 95.88 & 91.66 & 87.17 & 88.91 & 93.71 & 86.42
\\
MedicalNet~\cite{chen2019med3d} & 83.11 & 79.15 & 69.22 & 93.64 & 89.88 & 86.38 & 87.08 & 92.40 & 86.90
\\
Models Gen.~\cite{zhou2019models} & 82.97 & \textbf{83.05} & \textbf{75.49} & 95.79 & 92.90 & 90.10 & 91.06 & 94.65 & 85.78
\\
Swin UNETR~\cite{tang2022self}& \textbf{85.24} & 81.86 & 74.33 & 95.06 & 92.16 & 88.37 & 89.45 & 94.04 & 88.56
\\
UniMiSS~\cite{xie2022unimiss}& 82.11 & 79.37 & 73.12 & \textbf{96.08} & \textbf{93.18} & \textbf{90.31} & \textbf{91.99} & \textbf{95.43} & 88.51
\\
\hline
Universal Model & 85.21 & 82.25 & 75.01 & 95.04 & 92.28 & 88.21 & 89.69 & 94.06 & \textbf{88.95}
\\ \hline
\end{tabular}%
\label{tab:TotalSeg_organs}
\end{table*}

%%%%%%% JHH_finetuning
\begin{table*}[t]
\caption
{\textbf{The complete evaluation of JHH.} The results are evaluated by DSC. IVC (inferior vena cava), PSV (portal vein and splenic vein), AG (adrenal gland), CAA (celiac abdominal aorta)}
\centering
\footnotesize
\begin{tabular}{p{0.17\linewidth}|P{0.09\linewidth}P{0.09\linewidth}P{0.09\linewidth}P{0.09\linewidth}P{0.09\linewidth}P{0.09\linewidth}P{0.09\linewidth}P{0.09\linewidth}P{0.09\linewidth}}
\hline
\textbf{\textit{Method}}& spleen & Kidney\_R & Kidney\_L & gallbladder & liver & stomach &  
\\ \shline
Scratch & 95.66 & 94.43 & 93.69 & 86.14 & 96.74 & 94.30 & 
\\
MedicalNet~\cite{chen2019med3d} & 91.08 & 88.63 & 86.60 & 61.23 & 93.29 & 88.22 & 
\\
Models Gen.~\cite{zhou2019models} & 95.02 & 93.44 & 93.07 & 84.73 & 94.12 & 94.05 & 
\\
Swin UNETR~\cite{tang2022self} & 94.71 & 93.95 & 92.27 & 81.75 & 96.00 & 92.79 & 
\\
UniMiSS~\cite{xie2022unimiss} & 88.35 & 91.49 & 90.41 & 82.91 & 93.80 & 89.57 & 
\\
\hline
Universal Model & \textbf{95.98} & \textbf{94.71} & \textbf{94.00} & \textbf{87.18} & \textbf{96.87} & \textbf{94.50} &  
\\ \hline
\vspace{1.0 em}\\
\textbf{\textit{Method}} & aorta & IVC & pancreas & PSV & AG & CAA & Average
\\ \shline
Scratch & 87.68 & 79.73 & 85.03 & 68.48 & 66.61 & 50.61 & 81.98
\\
MedicalNet~\cite{chen2019med3d} & 83.27 & 75.32 & 70.67 & 46.82 & 41.69 & 26.87 & 68.88
\\
Models Gen.~\cite{zhou2019models} & \textbf{89.46} & \textbf{81.50} & 84.23 & \textbf{71.79} & \textbf{70.46} & \textbf{54.23} & \textbf{82.81}
\\
Swin UNETR~\cite{tang2022self} & 87.43 & 80.89 & 81.19 & 66.71 & 65.04 & 36.38 & 79.55
\\
UniMiSS~\cite{xie2022unimiss} & 88.50 & 77.98 & 71.86 & 61.68 & 51.82 & 49.16 & 76.10
\\
\hline
Universal Model & 88.36 & 79.98 & \textbf{85.82} & 69.38 & 65.88 & 50.53 & 82.24
\\ \hline
\end{tabular}%
\label{tab:JHH_finetuning}
\end{table*}

\section{Additional Evaluations}
\label{sec:additional_evaluation}

\tableautorefname~\ref{tab:msd_testset} shows the detailed numerical result between Universal Model and Swin UNETR. Tables~\ref{tab:TotalSeg_vertebrae}--\ref{tab:TotalSeg_organs} and \tableautorefname~\ref{tab:JHH_finetuning} show the per-class evaluation of TotalSegmentator and JHH, which validates the transferability of the proposed Universal Model.

\figureautorefname~\ref{fig:pseudo_truth_visualization} exhibits the contour line comparison among Universal Model and two human experts. We can see the model predictions are roughly similar to human annotation, which validates the effectiveness of the pseudo label generated by our Universal Model.

\figureautorefname~\ref{fig:qualitative_visualization_kidney_tumor} and \figureautorefname~\ref{fig:qualitative_visualization_liver_tumor} shows several kidney and liver tumor cases comparison among the proposed Universal Model and four competitive baseline methods. Our method can not only detect small and big tumors in various organs but also not generate false positives of tumors. 

\tableautorefname~\ref{tab:clip_ablation} shows the ablation study results of CLIP embedding, which is an extension for \tableautorefname~\ref{tab:clip_embedding}. Dice scores for each organ and tumor are reported.

\begin{figure*}[t]
\centerline{\includegraphics[width=1.0\linewidth]{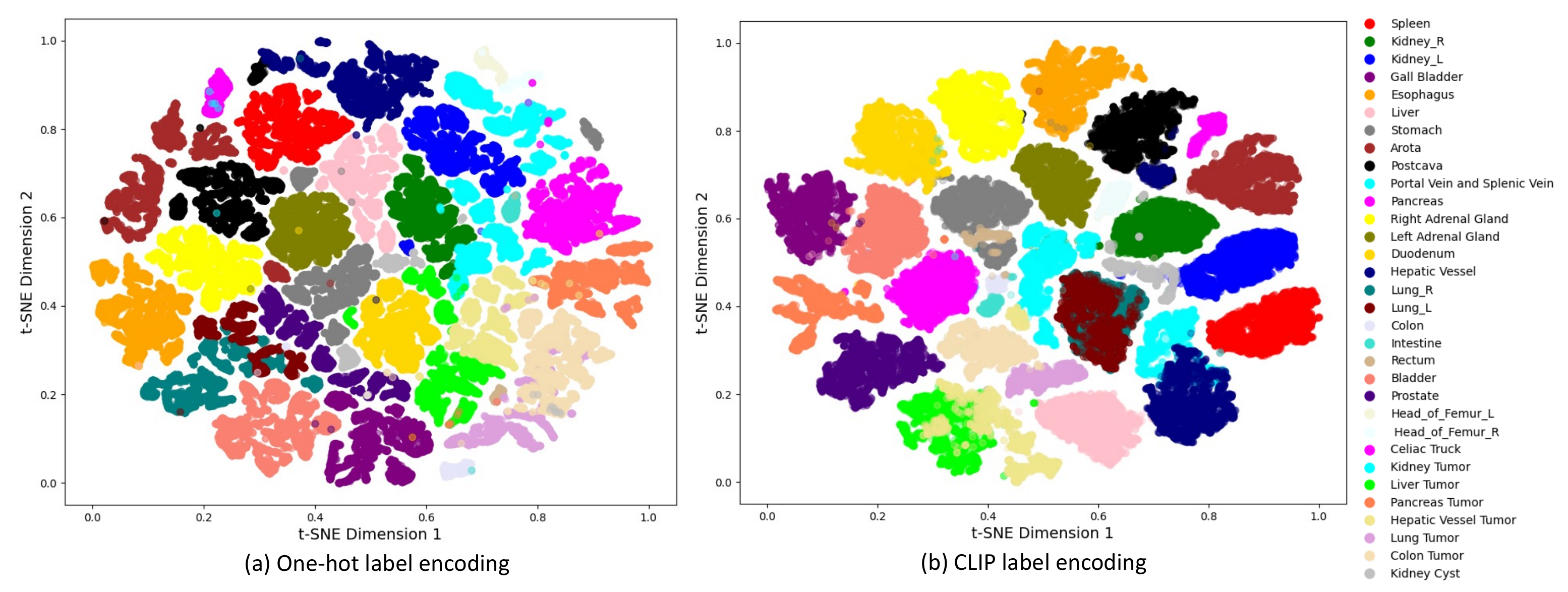}}
\caption{
\textbf{t-SNE Visualization of Whole Embedding Space.} Colors for corresponding embeddings are shown in figure.
}
\label{fig:all_tsne}
\end{figure*}

\figureautorefname~\ref{fig:all_tsne} shows the whole embedding space of baseline method and universal model. Our method shows better semantic relationship of anatomical structure.

%%%%%%% clip_ablation
\begin{table*}[t]
\caption
{\textbf{The complete results of embedding ablation study.} The results are evaluated by DSC. GB (Gallbladder), PSV (portal vein and splenic vein), AG (adrenal gland), HV (hepatic vessel), HF (head of femur), CT (celiac truck), KiT(kidney tumor), LiT (liver tumor), PT (pancreas tumor), HVT (hepatic vessel tumor), LuT (lung tumor), CoT (colon tumor), KiC (kideney cyst)}
\centering
\footnotesize
\begin{tabular}{p{0.13\linewidth}|P{0.05\linewidth}P{0.05\linewidth}P{0.05\linewidth}P{0.05\linewidth}P{0.05\linewidth}P{0.05\linewidth}P{0.05\linewidth}P{0.05\linewidth}P{0.05\linewidth}P{0.05\linewidth}P{0.05\linewidth}}
\hline
\textbf{\textit{Embedding}} & spleen & Kidney\_R & Kidney\_L & GB & Esophagus & Liver & Stomach & Aorta & Postcava & PSV & Pancreas
\\ \shline
One-hot~\cite{zhang2021dodnet} & 91.92 & 91.98 & 92.14 & 71.75 & 70.28 & 95.10 & 80.52 & 83.57 & 82.71 & 67.81 & 74.06
\\
BioBERT~\cite{yasunaga2022linkbert} & 94.65 & 93.26 & \textbf{92.98} & 75.14 & 72.32 & 95.09 & 87.68 & 91.05 & \textbf{83.91} & 67.83 & 80.51
\\
CLIP V1 & 92.35 & 91.83 & 91.89 & 72.45 & 71.38 & 90.23 & 73.07 & 86.77 & 78.17 & 74.00 & 74.91
\\
CLIP V2 & 93.05 & 92.14 & 91.42 & \textbf{75.88} & \textbf{75.56} & 94.75 & 75.79 & 91.15 & 80.64 & \textbf{78.90} & 78.94
\\
CLIP V3 & \textbf{94.69} & \textbf{94.09} & 92.77 & 73.45 & 72.87 & \textbf{95.71} & \textbf{89.19} & \textbf{92.19} & 83.44 & 59.20 & \textbf{86.09}
\\ \hline
\vspace{1.0 em}\\
\textbf{\textit{Embedding}} & AG\_R & AG\_L & Duodenum & HV & Lung\_R & Lung\_L & Colon & Intestine & Rectum & Bladder & Prostate
\\ \shline
One-hot~\cite{zhang2021dodnet} & 64.52 & 66.96 & 55.66 & 71.03 & \textbf{79.63} & 66.75 & 69.22 & 78.05 & 69.87 & 76.74 & 66.15
\\
BioBERT~\cite{yasunaga2022linkbert} & 65.94 & 68.72 & \textbf{68.61} & 59.14 & 75.40 & 69.09 & 71.24  & \textbf{81.78} & 65.58 & 74.51 & 69.51
\\
CLIP V1 & 72.07 & 72.42 & 62.42 & \textbf{74.53} & 79.32 & 76.52 & 70.32 & 75.65 & 63.11 & 75.06 & 66.47
\\
CLIP V2 & \textbf{79.98} & \textbf{79.73} & 66.01 & 68.65 & 75.87 & \textbf{82.98} & \textbf{74.88} & 70.82 & 64.64 & 70.06 & 68.8
\\
CLIP V3 & 64.75 & 70.18 & 71.11 & 65.43 & 77.48 & 62.11 & 71.77 & 81.47 & \textbf{79.42} & \textbf{86.71} & \textbf{72.96}
\\ \hline
\vspace{1.0 em}\\
\textbf{\textit{Embedding}} & HF\_L & HF\_R & CT & KiT & LiT & PT & HVT & LuT & CoT & KiC & Ave
\\ \shline
One-hot~\cite{zhang2021dodnet} & 70.27 & 60.23 & 78.92 & 63.84 & 68.02 & 55.48 & 52.31 & 53.87 & 48.39 & \textbf{35.81} & 70.42
\\
BioBERT~\cite{yasunaga2022linkbert} & 74.39 & 79.07 & 80.69 & 57.41 & 63.44 & 39.70 & 57.88 & 58.57 & 54.19 & 20.33 & 71.55
\\
CLIP V1 & 74.61 & 72.53 & 79.28 & 56.62 & 76.24 & 61.05 & 56.49 & 73.60 & 55.03 & 32.87 & 73.49
\\
CLIP V2 & 69.98 & 75.73 & \textbf{84.04} & 67.04 & \textbf{82.09} & \textbf{77.75} & 67.45 & \textbf{75.38} & 55.55 & 35.79 & 75.66
\\
CLIP V3 & \textbf{84.94} & \textbf{89.45} & 77.55 & \textbf{68.72} & 74.87 & 65.46 & \textbf{73.53} & 73.12 & \textbf{60.66} & 30.44 & \textbf{76.11}
\\ \hline
\end{tabular}%
\label{tab:clip_ablation}
\end{table*}

\begin{figure*}[t]
	\centering
	\includegraphics[width=1.0\linewidth]{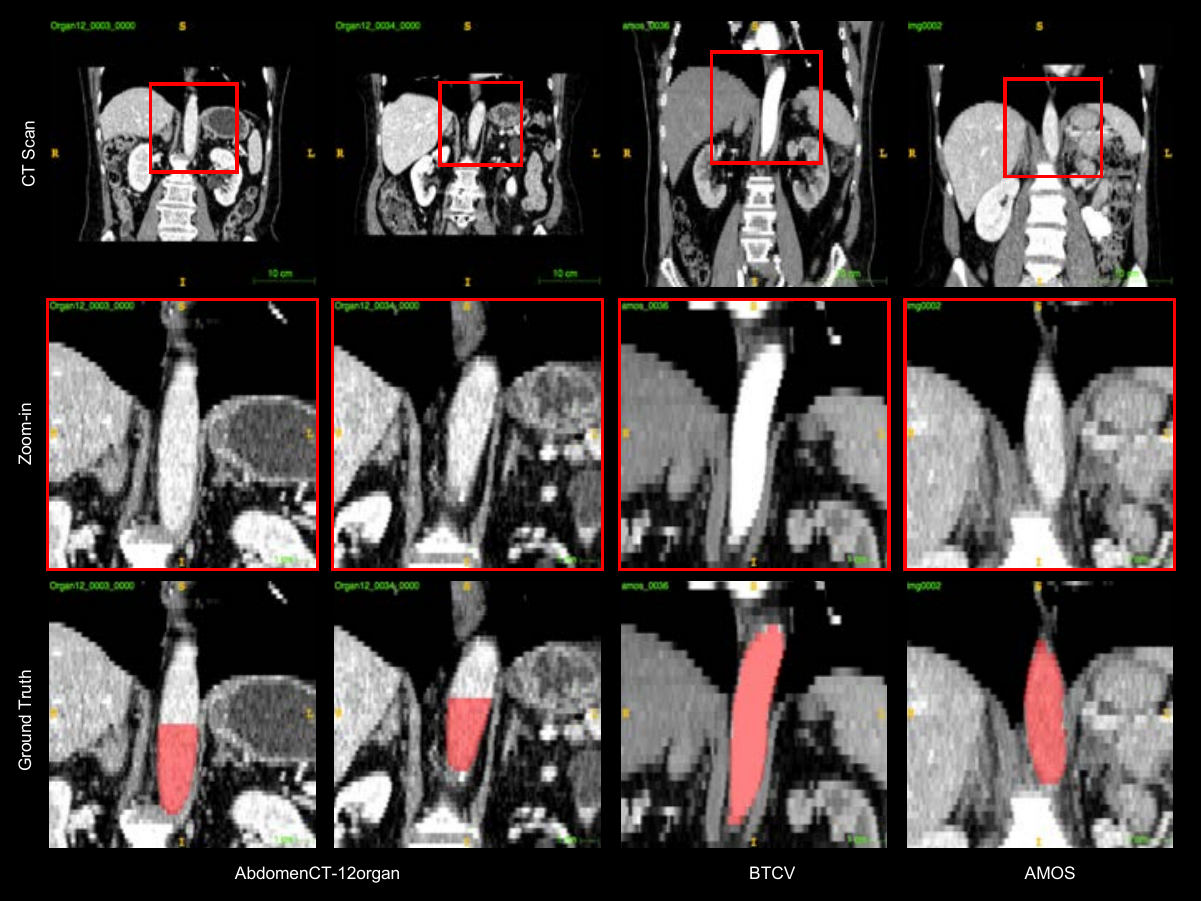}
	\caption{\textbf{Inconsistent Label Protocol.} The aorta annotation standard is inconsistent in AbdomenCT-12organ and other datasets. A part of the upper aorta region is missing in AbdomenCT-12organ, while the aorta annotation is complete in BTCV and AMOS.}
	\label{fig:inconsistent_protocol}
\end{figure*}

\begin{figure*}[t]
	\centering
	\includegraphics[width=\linewidth]{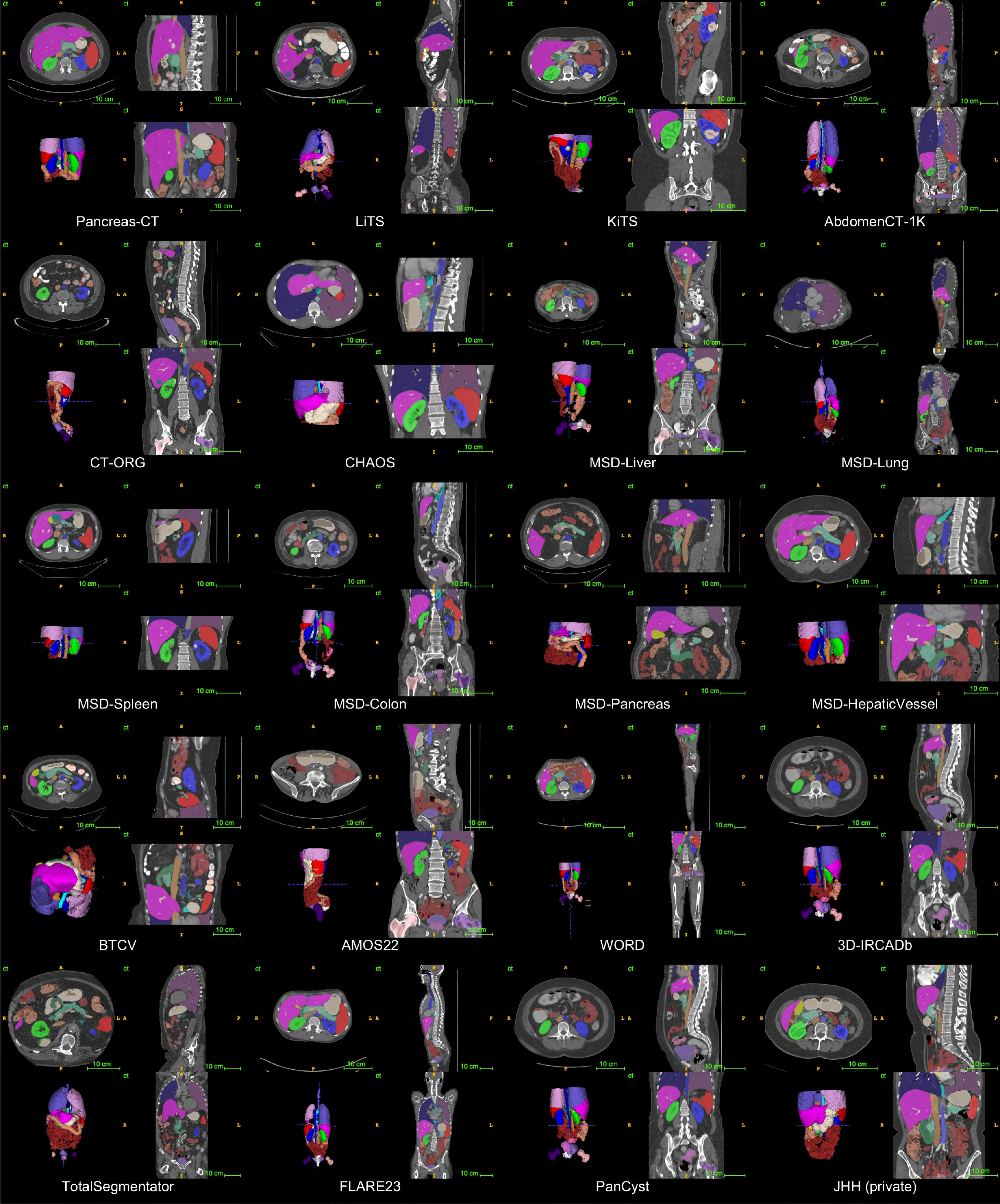}
	\caption{\textbf{Prediction of incomplete labels in previous datasets.} We leveraged the predictions generated by the Universal Model to produce masks for 25 organs in 20 CT datasets, achieving a satisfactory level of accuracy. However, we note that the accuracy of the 6-tumor segmentation still requires validation through pathology reports, which we have identified as a future direction for our work.}
	\label{fig:multiorgan_multitumor}
\end{figure*}

\begin{figure*}[t]
	\centering
	\includegraphics[width=0.75\linewidth]{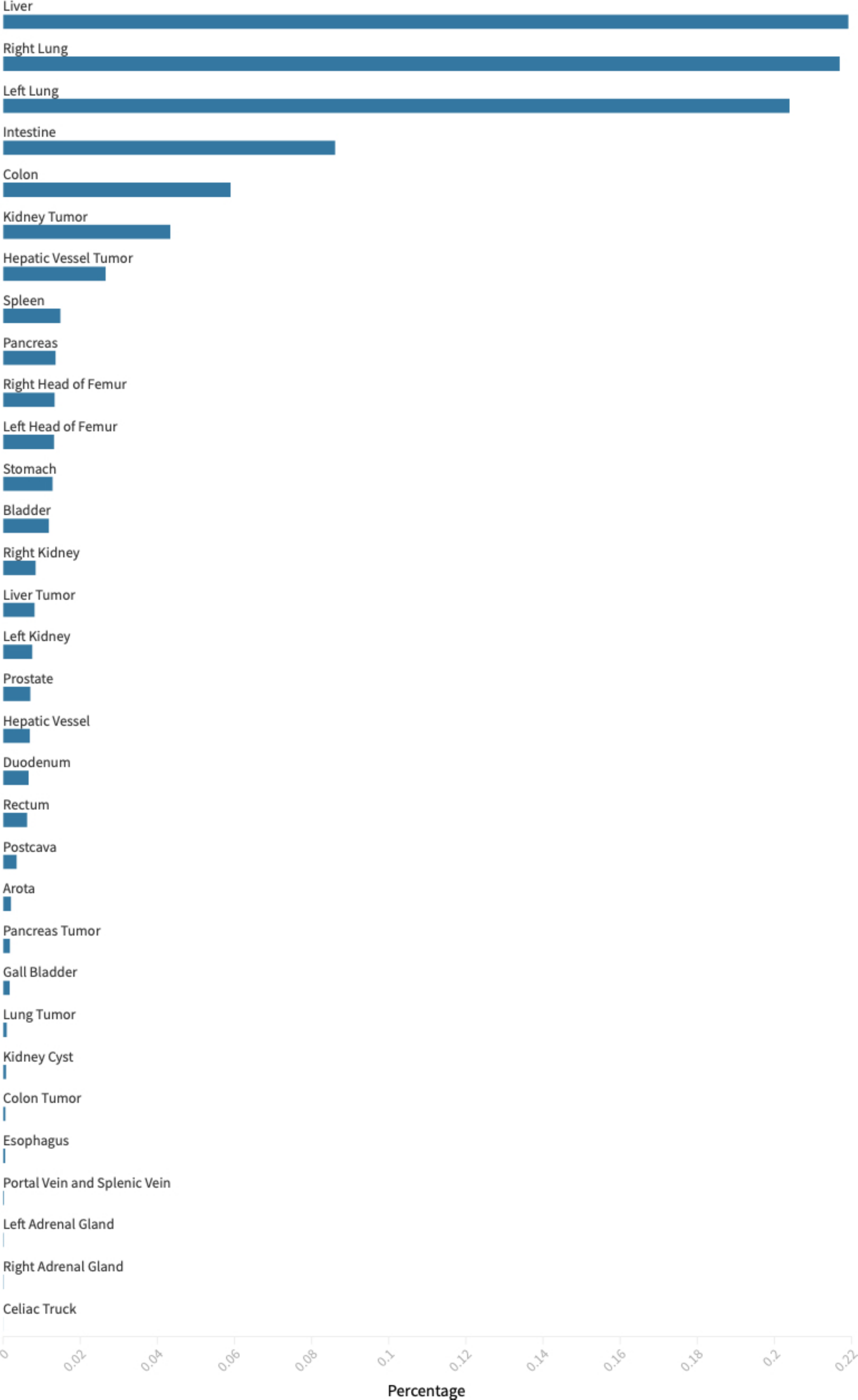}
	\caption{\textbf{The proportion of 32 classes.} We observe that the assembly of datasets presents severe long-tail distribution.}
	\label{fig:long_tail}
\end{figure*}

\section{Discussion of Open Challenges}
\label{sec:open_challenges_review}

\smallskip\noindent\textbf{\textit{Inconsistent label protocols.}} The first open challenge is the inconsistent annotation protocol. The annotation standard is different from institution to institution. In AMOS, ``Aorta'' refers to the entire region of Aorta, but in AbdomenCT-1K, a part of the upper regions annotation is missing. It is because of the inconsistent definitions in different datasets and this requires considerable manual corrections of several experienced radiology experts when assembling these datasets together. 

\smallskip\noindent\textbf{\textit{Long-tail problem.}} The assembly of public datasets leads to severe class imbalance problems, especially for small tumors. We count the proportion of each organ and tumor in \figureautorefname~\ref{fig:long_tail}. The assembly of datasets has a severe long-tail distribution, which would lead to unsatisfactory performance of tumor classes. Mitigating the long-tail distribution would contribute to more robust detection of the tumor. In this paper, we utilize data augmentation to alleviate the long-tail problem, but more research is encouraged to explore the solution to these two problems.

\end{document}